\documentclass[aps,pra,twocolumn,superscriptaddress]{revtex4-2}
\usepackage{graphicx}
\usepackage{amsmath,amssymb}
\usepackage{bm}
\usepackage[T1]{fontenc}
\usepackage{lmodern}
\usepackage{hyperref}
\usepackage{color}
\hypersetup{colorlinks=true,citecolor={blue},linkcolor={blue},urlcolor={blue}}
\usepackage{units}
\usepackage{soul}

\usepackage[normalem]{ulem}
\usepackage{xcolor}
\usepackage{upgreek}
\usepackage{pifont}

\makeatletter
\newcommand*{\rom}[1]{\expandafter\@slowromancap\romannumeral #1@}
\makeatother

\begin{document}

\title{Machine learning of phase transitions in nonlinear polariton lattices}

\date{\today}

\author{Daria Zvyagintseva}
\thanks{These authors contributed equally}
\affiliation{Computer Technologies Laboratory, ITMO University, St. Petersburg, 197101, Russia}
\author{Helgi Sigurdsson}
\thanks{These authors contributed equally}
\affiliation{Science Institute, University of Iceland, Dunhagi 3, IS-107, Reykjavik, Iceland}
\affiliation{School of Physics and Astronomy, University of Southampton, Southampton SO17 1BJ, United Kingdom}
\author{Valerii K. Kozin}
\affiliation{Science Institute, University of Iceland, Dunhagi 3, IS-107, Reykjavik, Iceland}
\affiliation{Department of Physics and Engineering, ITMO University, St. Petersburg, 197101, Russia}
\author{Ivan Iorsh}
\affiliation{Department of Physics and Engineering, ITMO University, St. Petersburg, 197101, Russia}
\author{Ivan A. Shelykh}
\affiliation{Science Institute, University of Iceland, Dunhagi 3, IS-107, Reykjavik, Iceland}
\affiliation{Department of Physics and Engineering, ITMO University, St. Petersburg, 197101, Russia}
\author{Vladimir Ulyantsev}
\affiliation{Computer Technologies Laboratory, ITMO University, St. Petersburg, 197101, Russia}
\author{Oleksandr Kyriienko}
\email{o.kyriienko@exeter.ac.uk}
\affiliation{Department of Physics and Astronomy, University of Exeter, Stocker Road, Exeter EX4 4QL, United Kingdom}

\date{\today}

\begin{abstract}
\textbf{Abstract:} Polaritonic lattices offer a unique testbed for studying nonlinear driven-dissipative physics. They show qualitative changes of a steady state as a function of system parameters, which resemble non-equilibrium phase transitions. Unlike their equilibrium counterparts, these transitions cannot be characterised by conventional statistical physics methods. Here, we study a lattice of square-arranged polariton condensates with nearest-neighbour coupling, and simulate the polarisation (pseudo-spin) dynamics of the polariton lattice, observing regions with distinct steady-state polarisation patterns. We classify these patterns using machine learning methods and determine the boundaries separating different regions. First, we use unsupervised data mining techniques to sketch the boundaries of phase transitions. We then apply learning by confusion, a neural network-based method for learning labels in the dataset, and extract the polaritonic phase diagram. Our work takes a step towards AI-enabled studies of polaritonic systems.
\end{abstract}

\maketitle

\section*{Introduction}
There is growing attention devoted to analysing physical systems through machine learning (ML) techniques given the ground-breaking advancements in artificial intelligence strategies~\cite{Carleo2019, Mehta2019}. With prominent examples of generative modelling \cite{Goodfellow2014}, recommendation systems \cite{Netflix2015}, natural language processing \cite{NLP2017}, decision processes and disease detection \cite{Lancet2019}, ML provides means to grasp data features that can escape the eyes of a trained professional. It has also initiated the effort in quantum ML to be performed by quantum devices \cite{Benedetti2019,Uvarov2020}. In the case of classification tasks, ML became a useful tool to reveal phase transition boundaries in spin systems~\cite{CarrasquillaMelko,Wang2016,Torlai2016,WHu2017,Nieuwenburg2017,KozinVK_ML_2019, Corte2020}, topological models~\cite{Deng2017,Rodriguez2019,TZWei2019,Canabarro2019,Balabanov2020,Scheurer2020}, photonic condensates~\cite{Rodrigues_PRL2021}, and strongly correlated fermionic systems \cite{Ohtsuki2016, Ohtsuki2017, Chng2017}. In quantum chemistry it is used to predict properties of organic compounds and perform high-throughput calculations \cite{Borysov2018,Olsthoorn2019}. In nanophotonics ML techniques are widely used for inverse design \cite{Piccinotti_2020,Wiecha2020}. Other examples include detection of Wigner function negativity in multimode quantum states~\cite{Cimini2020} and automatic learning of topological photonic phase transitions~\cite{kerr2020automatic,Lidiak2020}. In many cases ML gives greater insight into non-equilibrium systems~\cite{Cheng2020,Argonne2020} which are well known to host numerous nontrivial solutions~\cite{Cross_RMP1993}. Notably, many fundamental features in nature such as the complicated patterns appearing on animal coats~\cite{Turing_RS1952} and proliferation of defects in the Higgs field~\cite{Kibble_JPhysA1976} are linked to non-equilibrium analogues of phase transitions. 
This question was investigated in optical systems, specifically noting cooperative phenomena and self-organisation during lasing~\cite{Haken1975,Haken1978,HakenBook}.
The nature of such phase transitions was also studied in non-reciprocal systems~\cite{Fruchart_Nature2021} which describe systems with gain and loss. Similar physics can be studied in condensed matter systems, such as superfluids and Bose-Einstein condensates, offering an experimentally friendly strategy to explore such pattern formation and spontaneous self-organisation~\cite{Baumann_Nature2010} which can benefit from ML techniques.

Semiconductor microcavities~\cite{KavokinBook} in the strong light-matter coupling regime show increasing promise for studying novel nonlinear low-dimensional  optical phenomena. The normal modes in this regime are exciton-polaritons~\cite{CarusottoCiutiRev}, quasiparticles coherently composed of both excitonic resonances in embedded quantum wells and trapped photonic cavity modes. They enjoy the benefits of picosecond scale response times and high nonlinearity (particle interactions) coming from their photonic and excitonic parts, respectively. To date, various nonlinear effects were studied, showing polariton condensation (or lasing)~\cite{Kasprzak_Nature2006, Christopoulos_PRL2007, Schneider2013}, spin pattern formation~\cite{Leyder_NatPhys2007}, solitons~\cite{Chana2015}, vortices~\cite{Lagoudakis2008, Caputo2019}, quantum correlations~\cite{Delteil2019}, among many others~\cite{CarusottoCiutiRev}.

Perhaps the most exciting advancement are lattices of polariton condensates which have emerged as a promising way to create extended systems of trapped nonlinear light~\cite{Schneider_RepProgPhys2017}. They can be realised using a variety of techniques such a lithographically patterned inorganic~\cite{Jacqmin2014} and organic~\cite{Jayaprakash_ACS2020} cavities which act on the photonic mode, or using sculpted nonresonant lasers which act on the exciton mode~\cite{Wertz_NatPhys2010}. The latter case offers the interesting option of creating either ballistic gain guided~\cite{Pickup_NatComm2020,Topfer_Optica2020} or optically trapped~\cite{Askitopoulos2013, Ohadi2017, Ohadi_PRB2018} polariton condensates through the repulsive interactions between polaritons and photoexcited background excitons. Today, polariton lattices have enabled the studies for topological properties~\cite{Klembt2018, Kartashov2019, Sigurdsson_PRB2019, Liu_Science2020, Pickup_NatComm2020}, dispersionless bands~\cite{Whittaker2018, Goblot2019}, as analogue simulators of the XY-model~\cite{Berloff2017,Miri2020} and oscillatory networks~\cite{Kalinin_PRB2019}, and as optimisers for NP-hard problems~\cite{Kalinin_SciRep2018, Kalinin_PRL2018, Kyriienko_PRB2019}. 

With rapid improvements in the abovementioned techniques, the coherence length of polariton condensate lattices now greatly exceeds the typical unit cell size~\cite{Ohadi_PRB2018, Baboux_Optica2018,Topfer_Optica2020} which gives hope to study new and interesting phases of dissipative bosonic matter determined by the coherent flow of polaritons across the lattice sites. Indeed, in contrast to lattices, spatially uniform condensates are notoriously difficult to realise due to cavity disorder fragmenting the polariton fluid. Nonetheless, this idealised scenario has captured theoretical work in the recent years focused on dissipative Kibble-Zurek mechanisms through proliferation of vortices due to modulational instability~\cite{Liew_PRB2015}, spontaneous Turing patterns in resonantly driven systems~\cite{Werner_PRB2014}, nonequilibrium Berezinskii-Kosterlitz-Thouless phase transition in the optical parametric oscillator~\cite{Dagvadorj_PRX2015} and incoherent pumping~\cite{Zamore_PRL2020} regimes, and the critical exponent universality at long times~\cite{Comaron_PRL2018}. Formation of polarisation domain walls through the condensation (phase transition) quench~\cite{Solnyshkov_PRL2016} and XY spin phases~\cite{Wilson2016} were reported in lattice chains, and vortex street formation due to snaking instabilities in both resonantly~\cite{Koniakhin_PRL2019} and nonresonantly~\cite{Sigurdsson_PRB2017_parsol} driven polariton fluids. It is therefore of interest to develop and apply ML strategies for these driven-dissipative systems to facilitate understanding on how different phases are separated in this zoo of possibilities, especially in terms of the state-of-the-art condensate lattices.

In this paper, we use ML to classify phases of spinor exciton-polariton condensate lattices. We focus on recent experimental findings demonstrating highly nontrivial polarisation behaviour between optically trapped condensates resulting in both spontaneous and random pattern formation of the condensate polarisation (polariton pseudospin orientation)~\cite{Ohadi2017, Sigurdsson2017a}, a so-called spin-bifurcation regime. We have chosen this system since it offers a relatively simple experimental method to verify our findings through full Stokes polarimetry measurements on the emitted cavity light which carries information on the polariton pseudospin (or spin for short). We use ML to distinguish polarisation patterns across our lattice. This provides an efficient method to map out nonequilibrium phase boundaries. We sketch out the clustering of our multidimensional data and, using learning by confusion~\cite{Nieuwenburg2017}, we refine the boundaries between different phases. Our results are applicable to other observables across different driven-dissipative oscillatory systems such as coupled laser arrays and photonic condensates.
\begin{figure}
\includegraphics[width=\linewidth]{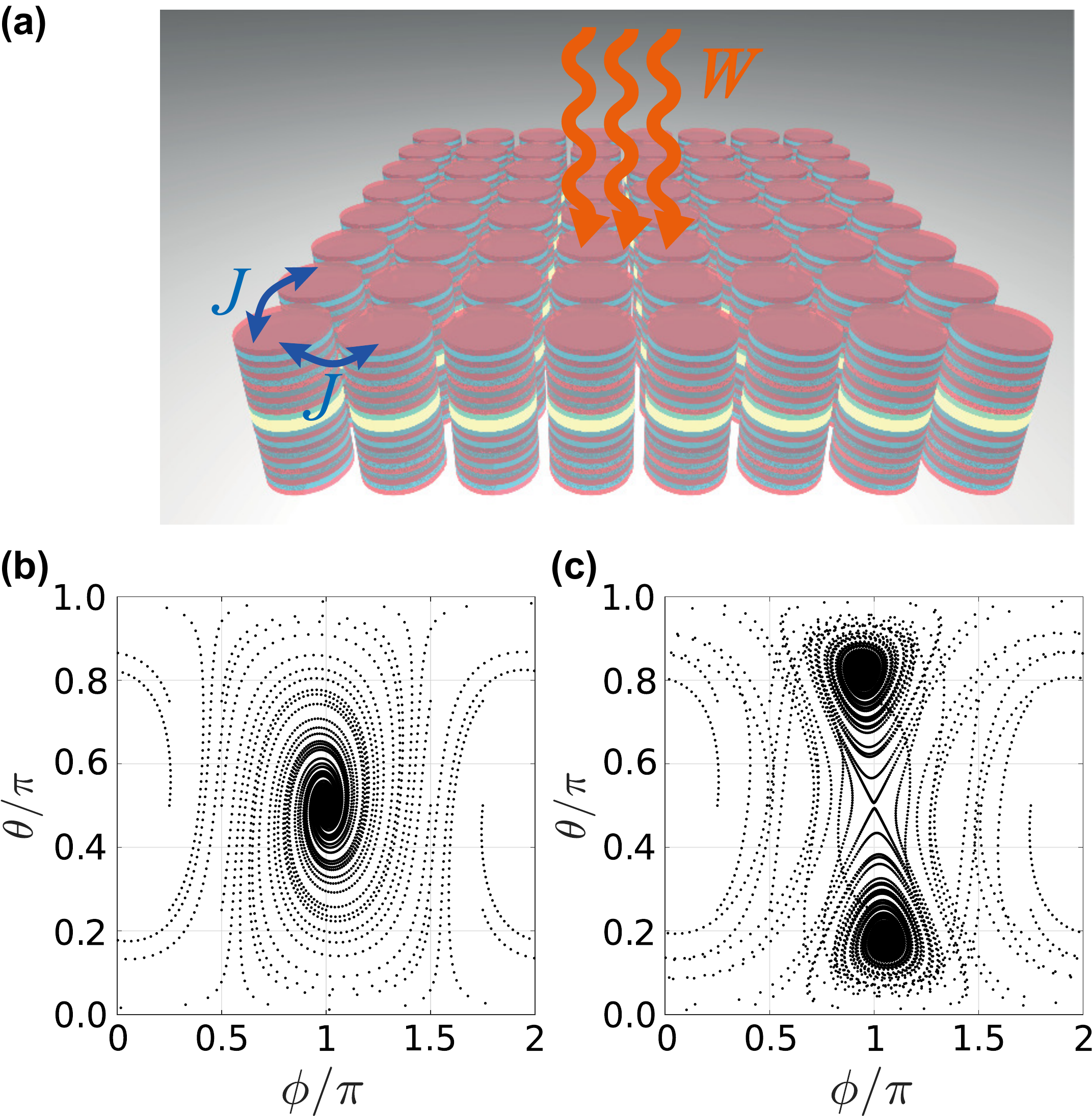}
\caption{\textbf{Lattice of coupled polariton condensates.} \textbf{(a)} Sketch of a square arranged polariton lattice based on coupled micropillars. $J$ denotes the tunnelling between sites and $W$ corresponds to the gain coming from the incoherent pump. \textbf{(b,c)} State space flow diagrams showing the evolution of the single condensate for several different initial conditions (here $\phi$ and $\theta$ are polar and azimuthal angles that parametrize pseudospin direction). We reveal the change from a single dominant fixed point attractor $s^z=0$ into two attractors of broken symmetry between spin-up and spin-down polaritons $s^z\neq0$.}
\label{fig_sketch}
\end{figure}

\section*{Results}

\subsection*{Model}

We consider a square lattice of optical cavities typically represented by coupled micropillars [see the sketch in Fig.~\ref{fig_sketch}(a)]. 
We consider the regime where the ground state mode of each pillar becomes macroscopically occupied by the polariton condensate. Each condensate is described by a coherent spinor wave function $\Psi_n = (\psi_{n+}, \psi_{n-})^\mathrm{T}$ for the $n$-th lattice site. The two spinor components $\psi_\pm$ correspond explicitly to the circular polarisation of the cavity light $\sigma^{\pm}$. The whole lattice is incoherently pumped by off-resonant linearly polarised light at high energy such that no phase or polarisation information is transferred from the laser source into the condensates. Such a system can be modelled using a set of coupled generalised spinor Gross-Pitaevskii equations~\cite{CarusottoCiutiRev},
\begin{align} \notag
         i \frac{d\Psi_n}{dt}  & =\frac{i}{2}
        \left(W_{t}(t) - \eta S_n \right)\Psi_n-\frac{1}{2}(\epsilon + i \gamma)\hat{\sigma}_x\Psi_n  \\ \label{eq.GP}
        & +\frac{1}{2}(\bar{\alpha} S_n + \alpha S^z_{n} \hat{\sigma}_z)\Psi_n -  (1-i\Lambda)\frac{J}{2} \sum_{\langle nm \rangle} \Psi_m,
\end{align}
where we have introduced the condensate pseudospin to describe the polarisation (magnetisation) of the lattice,
\begin{equation}
\mathbf{S}_n = (S_n^x,S_n^y,S_n^z)^\mathrm{T} = \frac{1}{2} \Psi^\dagger_n \boldsymbol{\hat{\sigma}} \Psi_n.
\end{equation}
Here $\boldsymbol{\hat{\sigma}} = (\hat{\sigma}_x, \hat{\sigma}_y, \hat{\sigma}_z)$ is the standard Pauli vector, and the magnitude of the spin for $n$-th condensate is $S_n = (|\psi_{n+}|^2 + |\psi_{n-}|^2)/2$. The factor $1/2$ is conventional. When presenting pseudospin patterns for the lattice we use normalised intensities at each site defined as $\mathbf{s}_n = \mathbf{S}_n/S_n$.
The parameters in the first line of Eq.~\eqref{eq.GP} include: $W_t(t)$ describing the time-dependent incoherent pump rate (gain) with subtracted linear losses (i.e., we have absorbed the conventional linear polariton loss parameter $\Gamma$, corresponding to the cavity photon escape rate, into our net gain parameter $W$); $\eta$ being a gain clamping (saturation) parameter describing isotropic nonlinear losses; $\epsilon$ and $\gamma$ being energy and linewidth (losses) splitting between the linearly polarised modes $\psi_{x,y} = (\psi_+ \pm \psi_-)/\sqrt{2}$. Physically, the complex valued linear polarisation splitting appears due to cavity strain~\cite{Ohadi2015}, leading to non-Hermitian coupling between circular polarisation components and defining the effective spin properties. The first term in the second line of Eq.~\eqref{eq.GP} describes the nonlinear shift of polariton energy due to polariton-polariton interactions for the same spin ($\alpha_1$) and opposite spin ($\alpha_2$) components. Specifically, in the circular polarisation basis we use the combinations $\alpha = \alpha_1 - \alpha_2$ and $\bar{\alpha} = \alpha_1 + \alpha_2$. Finally, the last term in Eq.~\eqref{eq.GP} describes the Josephson type coupling between lattice sites, $J$, and $\Lambda$ is an energy dampening parameter according to the Landau-Khalatnikov approach~\cite{Read_PRB2009}. The sum is to be taken over nearest lattice neighbours.

The system of equations \eqref{eq.GP} was found to describe successfully experiments on trapped polariton condensates~\cite{Ohadi2015, Dreismann_NatMat2016, Ohadi2017}. To study the condensates polarisation patterns, the incoherent pump is increased slowly and linearly in time until the target value $W$ is reached at the time $t_f$, 
\begin{equation} \label{eq.W}
W_t(t) = W \left( \Theta[t_f - t] \frac{t}{t_f} + \Theta[t - t_f] \right),
\end{equation}
where $\Theta[t]$ is a Heaviside step function. Starting from noisy background (stochastic initial conditions), the polaritons will condense (i.e., $S_n>0$ solution forms) when a critical threshold pump power $W_\text{cond}$ is reached. The condensation threshold is determined by the condition $S_n = 0$ and when a single eigenvalue of Eq.~\eqref{eq.GP} goes from having a negative imaginary part to positive imaginary part with increasing pump power $W_t$. This crossover takes place at $W_\text{cond} = -(\gamma + Z\Lambda J)$, where $Z=4$ is the number of nearest neighbours, and belongs to a linearly polarised solution written $S_n = - S_n^x$ (because $\gamma$ increases the gain for vertically polarised polaritons). We will throughout the paper refer to this linear polarisation regime as the XY phase in our ML analysis which refers to the fact that the pseudospin is lying on the equatorial plane of the Poincar\'{e} sphere. In the terms of amplitude oscillator models, the condensation point is also a bifurcation point marking the departure of the condensate (the oscillator) from the stable $S_n=0$ solution. We note that $W_\text{cond}<0$, which may seem counter intuitive from the perspective of ``negative power'', but arises naturally since our parameter $W$ describes the difference between pump gain and linear cavity losses.

When we further increase the pump power, the system becomes spontaneously circularly polarised at a second critical power value $W_\text{bif}$ even though the gain and saturation are spin isotropic and Eq.~\eqref{eq.GP} does not favour one spin projection over the other~\cite{Ohadi2015}. This phenomenon was labelled as a spin bifurcation. It allows for observation of spontaneous magnetic ordering between interacting condensates~\cite{Ohadi2017}, and can give rise to topologically protected elementary excitations~\cite{Sigurdsson_PRB2019}. Spin bifurcation can be demonstrated in the simplest case of a single condensate (i.e., $J=0$). Using the polariton pseudospin parametrized on the Poincar\'{e} sphere by the polar and azimuthal angles $\theta$ and $\phi$, we can express it as $\mathbf{s} = (\sin{\theta} \cos{\phi} , \sin{\theta} \sin{\phi}, \cos{\theta})^\mathrm{T}$. Solving the generalised Gross-Pitaevskii equation numerically for $W = 0$ and $W = 5/3$, and random initial conditions, we observe how the phase space flow transforms from one dominant fixed point attractor into two fixed point attractors just by increasing the pump [Fig.~\ref{fig_sketch}(b,c)]. 
This corresponds to spontaneous symmetry breaking for the $s^z$ spin projection, known as the polariton spin bifurcation~\cite{Ohadi2015}. The unit of time $t$ is taken in units of $\epsilon^{-1}$ and we used $\gamma = 0.2$, $\eta=\alpha_1 = 0.083$, $\alpha_2=-0.1\alpha_1$, and $\Lambda = 0.25$ similar to previous studies where the model was fitted to experimental observations~\cite{Ohadi2015}.

In order to determine the spin bifurcation pump power $W_\text{bif}$ we need to consider the stationary solutions of Eq.~\eqref{eq.GP} where each node has the same particle population $S_n = S_{n+1}$ and same magnitude spin polarisation $|S_{n}^z| = |S^z_{n+1}|$. It can be shown that solutions which satisfy the above requirements and minimise the bifurcation threshold are of the form~\cite{Sigurdsson2017a}
\begin{equation}
    \Psi_n = \left\{ 
  \begin{array}{l l}
\Psi_{n+1}, & \quad \text{if} \ \ S_n^z = S_{n+1}^z, \\
- \hat{\sigma}_x \Psi_{n+1}, & \quad \text{if} \ \  S_n^z = -S_{n+1}^z. \\
\end{array} \right.
\end{equation}
These trivial solutions characterise ferromagnetic and antiferromagnetic states where two condensates are spin parallel with zero phase slip between them or spin antiparallel with a $\pi$ phase slip between them respectively. The bifurcation threshold is dictated by the parameters of the system and possible spin arrangement between nearest neighbours,
\begin{equation} \label{eq.bif}
W_\text{bif} =  W_\text{cond} + \eta \frac{(\epsilon - Z_{\uparrow \downarrow} J)^2 + (\gamma + Z_{\uparrow \downarrow} \Lambda)^2}{\alpha (\epsilon - Z_{\uparrow \downarrow} J) } .
\end{equation}
Here, $Z_{\uparrow \uparrow}$ and $Z_{\uparrow \downarrow}$ are the number of nearest neighbour ferromagnetic and antiferromagnetic bonds for a condensate in the lattice (equal for all nodes). In general, Eq.~\eqref{eq.bif} states that a stationary polarisation pattern of certain parallel and antiparallel nearest neighbour spins may arise when $W_t$ is increased to $W_\text{bif}$. However, it is not known beforehand what determines the exact outcome of Eq.~\eqref{eq.GP} starting from some initial state vector. For example,  $Z_{\uparrow \uparrow} = Z_{\uparrow \downarrow} = 2$ patterns have many different possible configurations for a given lattice size which all have the same bifurcation point $W_\text{bif}$. We also do not know the stability of these steady state solutions and what other solutions might exist. Apart from ferromagnetic and antiferromagnetic bonding configurations between nearest neighbour condensates one can expect more complex states to appear which can be categorised broadly as stationary, cyclic, and chaotic with condensate patterns of varying spin and magnitude. Our goal is to use ML to characterise and cluster these patterns.



Next, we continue to present our numerical results. Specifically, we describe: 1) the numerical procedure of generating the dataset of polariton polarisation patterns; 2) details of the data analysis and visualisation; 3) mapping of coarse-grained phase boundaries and qualitative description of the zoo of phases; 4) introduce unsupervised ML methods; 5) and present the phase diagram of the polariton lattice spin phases.


\subsection*{Numerical simulations}
We consider an $8 \times 8$ polariton lattice and numerically solve generalised Gross-Pitaevskii equations (see details in the Numerical modelling subsection in Methods).
\begin{figure}
    \centering
    \includegraphics[width=\linewidth]{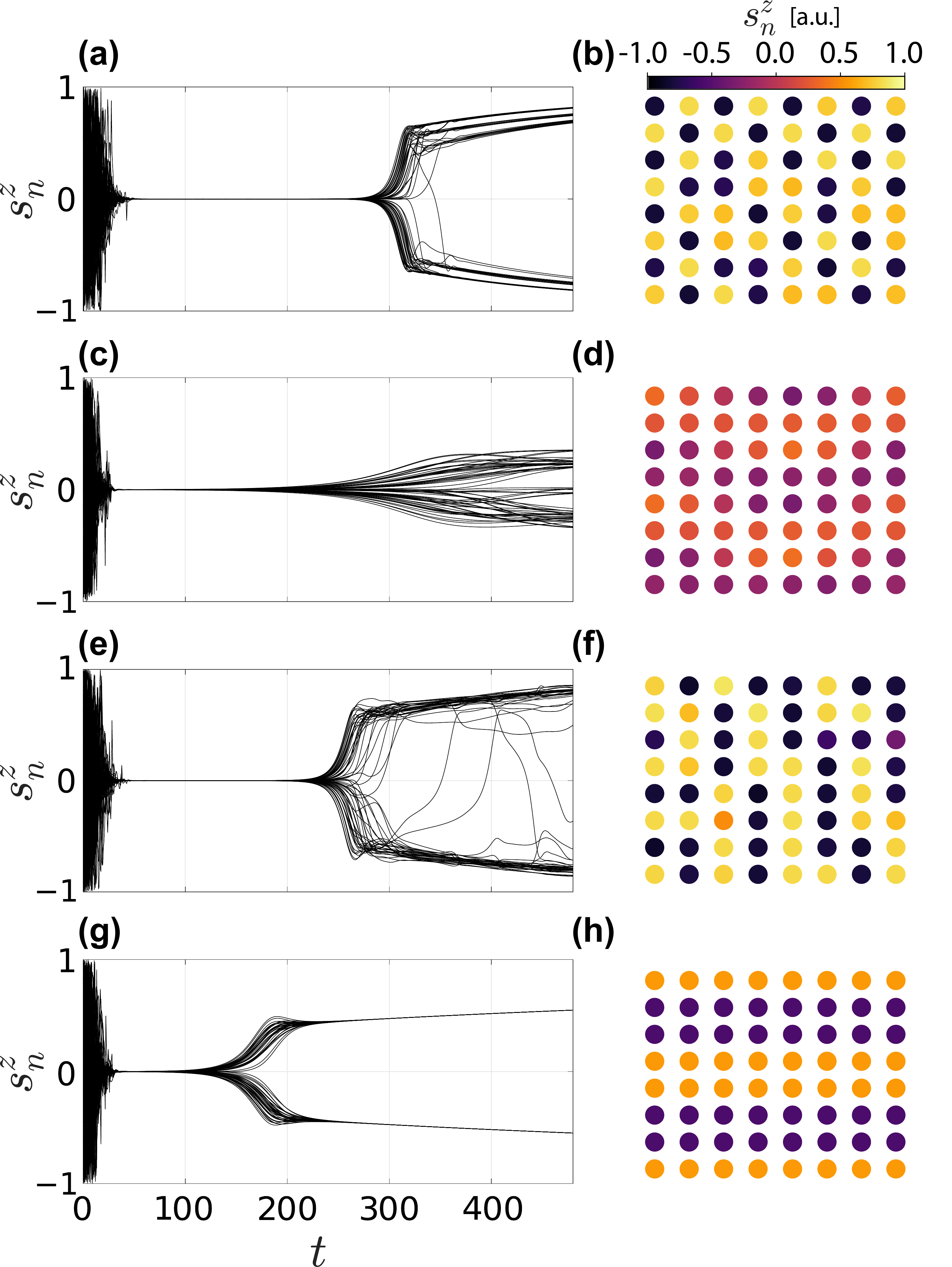}
    \caption{\textbf{Polariton lattice dynamics.}
   In the left column we show examples of dynamical trajectories $s_n^z(t)$ for a $8\times8$ lattice of condensates for different values of $J$ and $W$. Overlaid black lines correspond to different condensates in the lattice. In the right column we show the corresponding normalised magnetisation $s_n^z(t_f)$ at final time $t_f = 480$ (measured in dimensionless units labelled as a.u.). Depending on $W$ and $J$ distinct polarisation patterns appear with hints of the antiferromagnetic order \textbf{(a,b)}, weak circular polarisation \textbf{(c,d)}, two spin-down and two spin-up neighbours \textbf{(e,f)}, and the striped pattern \textbf{(g,h)}. Note the strong non-convergent character of the dynamics in \textbf{(e)}.}
    \label{fig:trajectories}
\end{figure}
In Fig.~\ref{fig:trajectories} we show an example of four simulations of the full lattice polarisation. In Figs.~\ref{fig:trajectories}(a,c,e,g) we plot normalised $s^z_n(t)$ spin components is for all sites as a function of time. In Figs.~\ref{fig:trajectories}(b,d,h,f) we plot final polarisation patterns measured at $t_f$, where colour bars encode the magnitude of the spin component $s^z_n(t_f)$. The four examples shown in Fig.~\ref{fig:trajectories} are picked from a set of 100 unique simulations with random gain $W$ and coupling strength $J$ to illustrate the plethora of phases appearing in our system. Specifically,  Figs.~\ref{fig:trajectories}(a,c,e,g) correspond to $W = \{0.77, 0.005, 0.69, 0.12 \}$ and $J = \{0.13, 0.48, 0.24, 0.48 \}$, respectively. To model experimental conditions, we also use stochastic initial conditions. 

The resulting dynamics can correspond to both stationary [Figs.~\ref{fig:trajectories}(b,d,h)] and nonstationary patterns [Figs.~\ref{fig:trajectories}(f)]. The latter emerge due to the interplay of drive, decay, and nonlinearity in the system. 
Our goal is to find stationary states with distinct polarisation pattern formation that can be seen as phases of matter for polaritons, which we refer as polaritonic phases in the following. We observe that various polaritonic phases can emerge as analogues of spin phases, albeit in the driven-dissipative setting. For instance, in Fig.~\ref{fig:trajectories}(b) we observe a spin pattern that resembles antiferromagnetic ordering with $[Z_{\uparrow \uparrow}, Z_{\uparrow \downarrow} = (0,4)]$. At the same time, several different patterns were observed, such as the paired neighbour phase $[Z_{\uparrow \uparrow}, Z_{\uparrow \downarrow} =(2,2)]$ in Fig.~\ref{fig:trajectories}(f), which have remained unexplored.

Having observed qualitatively different behaviour for polarisation of the nonlinear polaritonic lattice, we may ask a question: how do we classify and draw boundaries between different polaritonic spin phases? Unlike the thermodynamic equilibrium case, in the driven-dissipative case we do not have an established theory of phase transitions~\cite{Wilson2016}. We therefore take a data-driven approach, and use ML for unsupervised clustering of polaritonic phases.
\begin{figure*}
\includegraphics[width=1.\linewidth]{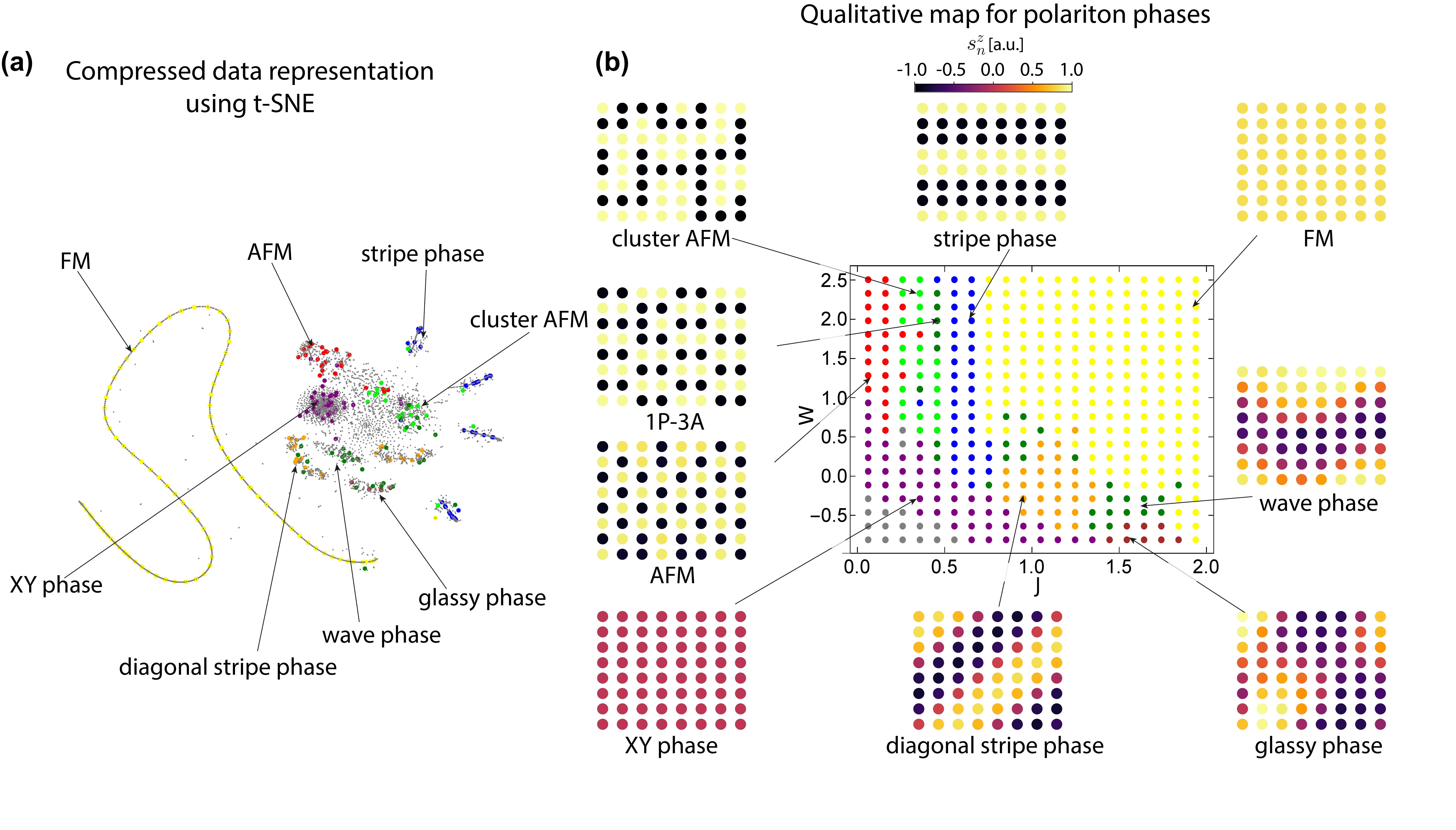}
\caption{\textbf{Visualisation of tentative polariton phases}. \textbf{(a)} Data visualisation of polarisation patters using the dimensionality reduction method. Results are obtained using the t-distributed stochastic neighbour embedding (t-SNE) approach on the raw dataset with $\{ \mathbf{s}_n \}$. We observe qualitative clustering of characteristic polarisation patterns, and label them by checking the lattice magnetisation for marked points (solid dots). \textbf{(b)} Qualitative map of polariton phases, shown coloured dots in the pump-tunnelling coordinates (central plot). The map is extracted from t-SNE data and the performed pattern analysis. Hypothetical classes are shown by coloured dots (see the legend below), and we provided typical instances of the lattice polarisation $s_n^z$ (insets), where the top colour bar is the same for all lattices. The labels correspond to: AFM -- antiferromagnet (red dots), FM -- ferromagnet (yellow dots), 1P-3A -- one parallel three antiparallel configuration (dark green dots), cluster AFM (green dots), stripe phase (blue dots), XY phase (purple dots). Wave, glassy, and diagonal stripe phase are shown by   and others being self-explanatory. The grey region (bottom left) corresponds to lattices where polaritons are not condensed (i.e., $S_n = 0$).}
\label{fig:tSNE}
\end{figure*}

\subsection*{Data visualisation} The prepared dataset of polarisation patterns contains $\{\mathbf{S}_n\}$ lists with $192$ entries for each point on the equally-spaced grid $\{ J_j, W_k \}_{j,k}$. We set $t_f = 3000$ in Eq.~\eqref{eq.W} such that all other timescales are surpassed. While the full dynamics is obtained by numerical propagation of Eq.~\eqref{eq.GP}, in practice we only retain data at the last timesteps at $\mathbb{T} = \{ t_f + i \delta t\}_{i=0}^{10}$ with $\delta t = 1$. Next, we perform pre-processing on the raw data to ensure only relevant configurations are studied. For this, we discard nonstationary data points where the variance (difference) between spin patterns in the time series $\mathbb{T}$ is greater than some sensibly chosen tolerance, and concentrate only on stationary states. For convenience, we also filter out redundant configurations differing only through trivial symmetry operations in the sign of $s^z_n$ (for example, the two types of lattices in an antiferromagnetic arrangement). This is done by performing a rotation, which corresponds to changing the signs of $s^y_n$ and $s^z_n$ pseudospin components in cases where $s^z_n$ have the same magnitude.

We then proceed by analysing the high dimensional data. The starting point corresponds to data visualisation through the dimensionality reduction. We employ two methods corresponding to the t-distributed stochastic neighbour embedding (t-SNE)~\cite{vanderMaaten2008} and principal component analysis (PCA). These techniques allow for plotting datasets in a low-dimensional feature space (two or three dimensions). 

Performing PCA for the dataset we can potentially identify the most important features of the condensate spin lattice. Namely, PCA converts data points into a set of sequential orthogonal components and maximises the magnitude of the sample variance. This can be used as an additional pre-processing step before t-SNE analysis (choosing most relevant features), or for two- and three-dimensional visualisation. For the specific problem we consider, however, PCA did not prove useful for the visualisation of the polaritonic dataset. Complex polarisation patterns cannot be easily distinguished by the dominant principal component (e.g. total magnetisation $M_z = \sum_n S_n^z$). This prompts us to use t-distributed stochastic neighbour embedding instead.

We use t-SNE as a tool for finding points in the parameter space that share similar behaviour (see details in the Visualisation subsection in Methods). In the reduced space t-SNE locates points in a way that similar patterns are placed together, while distinct patterns are shown by distant points (with high-probability). This property is useful for mapping the hypothetical phase boundaries, where t-SNE offers a visualisation for clusters of points with qualitatively similar behaviour. We note however that t-SNE does not preserve the distance between points, and can only help drawing qualitative conclusions. 
\begin{figure}
\includegraphics[width=1.0\linewidth]{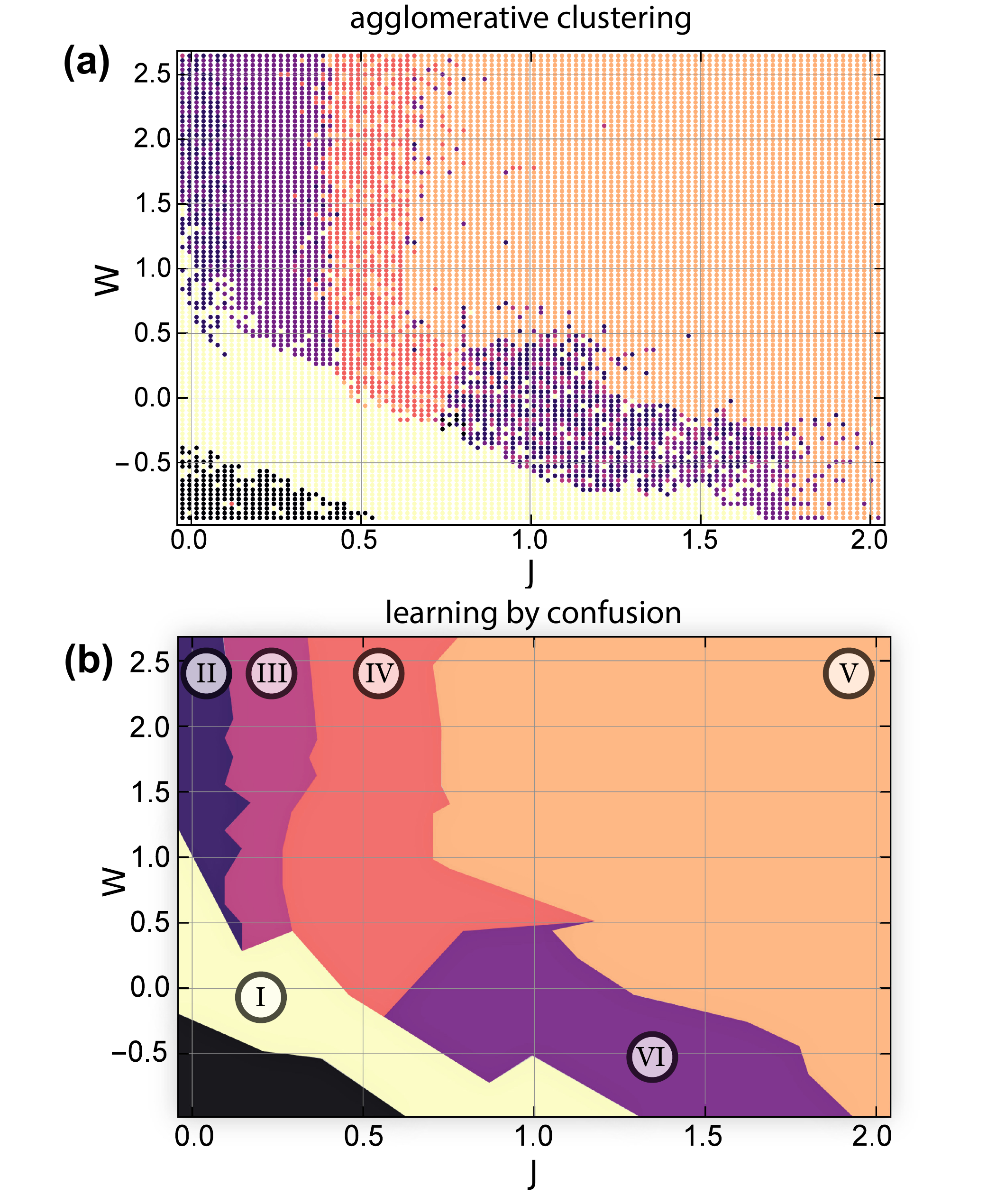}
\caption{\textbf{Polariton phase diagrams.} \textbf{(a)} Polariton phases separated using the agglomerative clustering. The diagram is built with the Manhattan metric, complete distance, for $s_n^x, s_n^y, s_n^z$ components and principal component analysis with 5 principal components. \textbf{(b)} Polariton phase boundaries obtained using a learning by confusion algorithm. We separated six distinct regions (labelled by \rom{1}--\rom{6}) in the pump vs tunnelling rate coordinates. Comparing results with the qualitative map we confirm the presence of: \rom{1} -- XY ordering, \rom{2} -- antiferromagnet (AFM) phase, \rom{3} -- clustered AFM, \rom{4} -- stripe ordering, \rom{5} -- ferromagnet (FM) phase, \rom{6} -- diagonal stripe ordering. The black region shows the range of parameters below the condensation point.}
\label{fig:agglomerative}
\end{figure}

In Fig.~\ref{fig:tSNE}(a) we show the two-dimensional t-SNE data visualisation for the data set with $s_n^x$ and $s_n^z$ components. Specifically, we use the medium perplexity level of 100 and the learning rate of 200. We note that the resulting t-SNE diagram does not change qualitatively with the change of hyperparameters, and similar results can be obtained in the broad range of perplexities and learning rates. Each sample is represented by a thin grey dot. Additionally, we check the polarisation patters (examples) from a sparse set of $\{ J_j, W_k \}_{j,k}$ values, shown as thick dots in Fig.~\ref{fig:tSNE}(a). Qualitatively similar patterns are drawn in the same colour. In Fig.~\ref{fig:tSNE}(b) we present these examples of polaritonic phases, forming the qualitative map and giving them tentative names.  Specifically, we identified: an XY phase where $\mathbf{s}_n = (-1,0,0)^\text{T}$; chequerboard antiferromagnetic patterns corresponding to the 2D antiferromagnet (AFM); cluster AFM patterns with zero total $z$-magnetisation, and configurations where 2 nearest neighbours are spin-aligned, and 2 nearest-neighbours are anti-aligned, $[Z_{\uparrow \uparrow}, Z_{\uparrow \downarrow} =(2,2)]$; stripe phase with zero total z-magnetisation and $[Z_{\uparrow \uparrow}, Z_{\uparrow \downarrow} =(3,1)]$; a ferromagnetic phase with uniform spin values of $s^z_n \approx \pm 1$. We find that configurations with $[Z_{\uparrow \uparrow}, Z_{\uparrow \downarrow} =(1,3)]$ (1P-3A) are rare and generally unstable. Additionally, we observe patterns with non-homogeneous polarisation distributions. We label them as: a hypothetical wave phase (similar patterns occupying high $J$ and intermediate $W$ region); a glassy phase with emergent domains of reverted polarisation on the dominant background; a diagonal stripe phase with continuous change of $s^z_n$ along the diagonals (distinct from the horizontal/vertical stripe phase).

This zoo of discussed polaritonic phases serves us as a base hypothesis. The question is: do we indeed label distinct driven-dissipative phases defined by unique polarisation patterns in the condensate lattice, or are these simply mixed and frozen patterns between conventional FM and AFM configurations? Next, we test the hypothesis using the unsupervised clustering and neural network (NN) based learning by confusion approaches.


\subsection*{Unsupervised learning} 

We now have a map of the polariton condensate lattice phases. Our next step is to perform unsupervised clustering. This procedure analyses the underlying data structure of an unlabelled dataset. The goal is to provide labels for data points, separating them into distinct groups. These groups share similar properties, in our case being stable and stationary spin patterns. We remind that each data point (associated with specific $J$ and $W$) corresponds to a high-dimensional vector $\bm{v}$ describing raw polarisation components $\{ \mathbf{s}_n \}$ or compressed feature vectors $\{ \mathbf{p}_i \}$. 

In the polariton dataset analysis we use the agglomerative and $K$-means clustering realisations from {\sffamily{}sklearn} library (see details in the Clustering subsection in Methods). The clustering algorithms are applied to both the raw dataset and the pre-processed dataset with a chosen number of principal components. Other possible choices concern the selection of metric and distance types. To choose a setting of high-performance, we develop a quality score, where good choices consistently assign same labels to data points in the three well known phases (XY, AFM, FM). 
We achieve best result for $\{\mathbf{s}_n\}$ data pre-processed with PCA and considering five principal components. We identified the optimal distance choice as the complete distance with the Manhattan metric.
Applying the agglomerative clustering procedure and labelling each data point, associated to one cluster, by different colours, we plot the resulting phase diagram in Fig.~\ref{fig:agglomerative}(a). Comparison with the qualitative map inferred from t-SNE [Fig.~\ref{fig:tSNE}(b)] allows us to assess the quality of clustering. We observe the phase boundaries in certain parameter regions. In particular, between XY phase, antiferromagnetic ordering, and ferromagnetic ordering are visible. At the same time, while we see that several qualitatively different antiferromagnetic patterns appear at small $J$ and high pump $W$, the boundaries within are difficult to establish. Finally, the region of $0.75<J<1.5$ and $-0.5<W<0.5$ with diagonal stripes and spin-glass patterns does cluster out, but contains varying labels that correspond to those with antiferromagnetic orderings.
Performing $K$-means clustering, we observe qualitatively the same performance for $K=6$, thus suggesting that some of previously identified phases cannot be categorised as such. To get more quantitative insight into the polariton lattice physics, we apply NN-based methods and further test and refine the phase boundaries.
\begin{figure}
\includegraphics[width=1.0\linewidth]{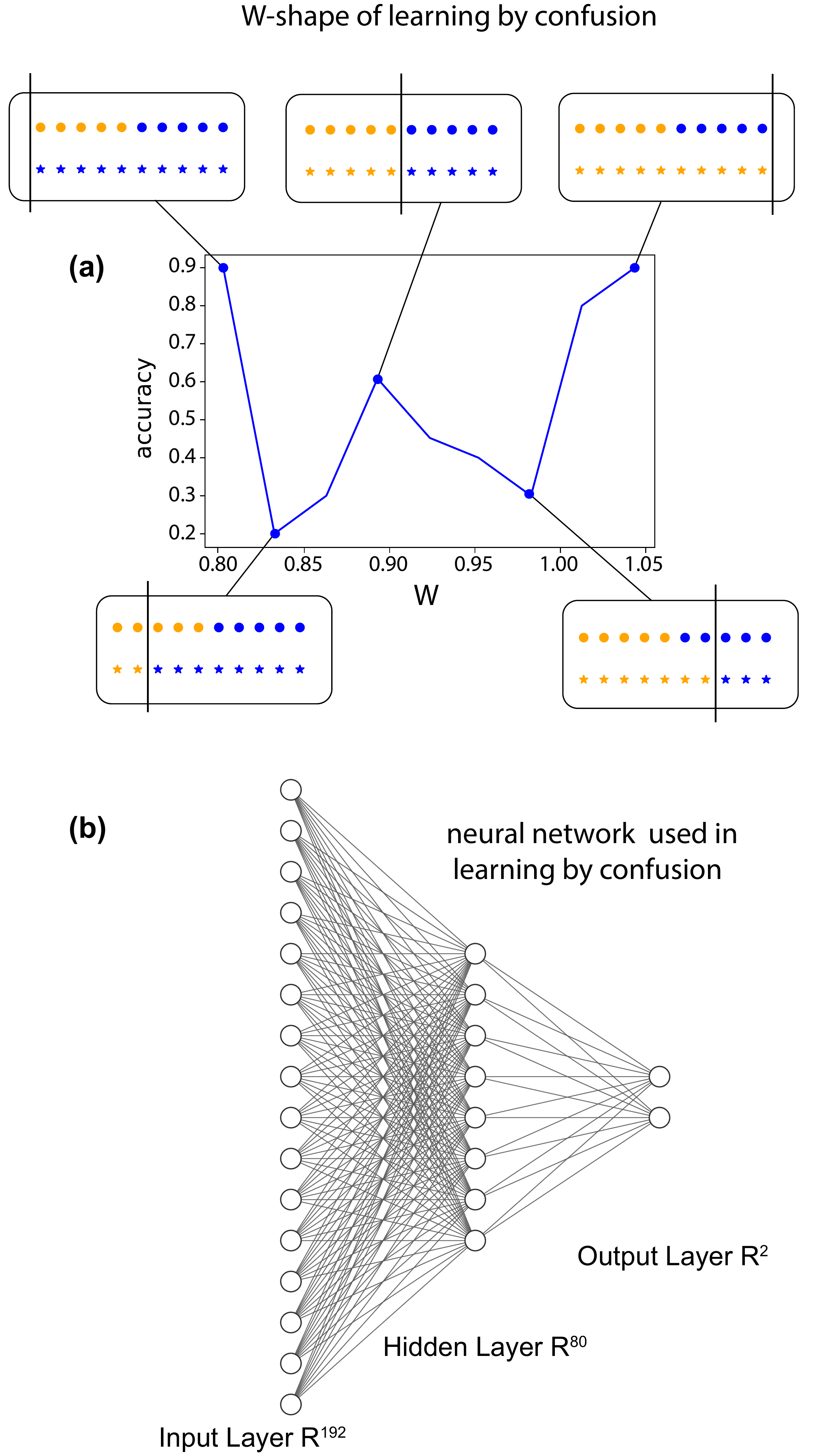}
\caption{\textbf{Learning by confusion.} \textbf{(a)} An example W-shape of the accuracy of neural network training during learning by confusion. We fixed the tunnelling rate $J$ and varied the lattice gain parameter $W$, observing peak accuracy in training when the hypothetical labelling coincided with the genuine one. The insets show cartoons for possible types of labelling. Circles show genuine labelling corresponding to two phases (yellow and blue), with the true critical point placed in the middle. The hypothetical labelling is shown by stars. \textbf{(b)} The structure of the neural network used in learning by confusion. It contains three layers: input layer ($64 \times 3$ neurons), hidden dense layer (80 neurons) and output dense layer (2 outputs).}
\label{fig:LbC}
\end{figure}

\subsection*{Learning by confusion}
While unsupervised learning methods allow to screen datasets and mine qualitative results, typically they are not suitable to determine phase boundaries. In contrast, supervised learning has shown great potential in determining phase boundaries using the power of NNs \cite{CarrasquillaMelko,Carleo2019}. They assume that the representative candidates for the phases are known, for instance, defined by zero and infinite temperature limits in classical spin systems. Typically, the datasets of spin patterns are formed by Monte Carlo procedures, where each point in the parameter space (temperature, interactions, etc) is assigned to a collection of similar patterns. Training the NN as a classifier then allows for identifying the boundary between distinct collections (or phases in the physical sense). In the absence of prior labelling and multiple phases the direct application of supervised training is infeasible.
In the following, we use NN-based technique that allows us to determine the phase boundaries without prior knowledge of the phases (no phase labels are provided). This corresponds to the learning by confusion (LbC) approach proposed in Ref.~[\onlinecite{Nieuwenburg2017}].

The main idea of LbC is in providing hypothetical labelling and then using supervised training to identify regions where the hypothesis is justified. For simplicity, we will discuss a one-dimensional phase boundary determination, where one of the parameters is fixed. A full phase diagram is then obtained by consecutive line-by-line scanning (both $J$ and $W$ can be fixed and scanned interchangeably). First, let us describe the details of the LbC approach. For this, consider a system that shows a qualitative different behaviour as a function of the parameter $W$. This corresponds to two phases separated by the critical point located at a certain (unknown) pump power $W_{\mathrm{crit}}$ [see the sketch in Fig.~\ref{fig:LbC}(a), insets]. To infer the critical point we can train a NN assigning hypothetical (fictitious) labels, where a candidate for the critical point $W_0$ is chosen on the interval from $W_1$ to $W_2$. All points for $W < W_0$ are considered to be in the first phase (labelled as ``yellow''), and points for $W > W_0$ are in the second phase (labelled as ``blue''). This corresponds to our hypothesis that needs to be tested for a set of candidate critical points. Note that labelling is applied both to training and testing sets used in variational NN optimisation.
We start by setting the critical point to be at the end of the interval, $W_0 = W_1$. In this case all data points are assigned to the group ``blue''. Next, we test the accuracy of the trained network defined as the probability in which predictions match the provided labels. We obtain 90\% accuracy, as test data also contains only examples with a single label. Same situation holds at the other end of the interval, $W_0 = W_2$. However, the situation changes when $W_0$ is placed between $W_1$ and $W_2$, and two labels are present. In this case, unless $W_0$ corresponds to the true critical point $W_{\mathrm{crit}}$, we are training the network to put qualitatively different data points (feature vectors) in the same phase, leading to confusion and reduced accuracy. The accuracy approaches unity when labelling is performed correctly, meaning $W_0 = W_{\mathrm{crit}}$. This happens because the inner structure of the lattice matches the markup. The overall behaviour for the accuracy thus resembles a W-shape~\cite{Nieuwenburg2017} (not to be associated with the parameter $W$), and is symmetric if $W_\mathrm{crit}$ is located in the middle of the $[W_1, W_2]$ interval. In other words, the point of phase transition corresponds to the point where the first derivative of the described accuracy function changes sign from plus to minus.

To perform LbC we construct a feed-forward NN with three layers [see the NN structure in Fig.~\ref{fig:LbC}(b)]. The first input layer consists of $192$ neurons such that the raw data $\{ \mathbf{s}_n \}$ can be analysed. The input leads to the fully-connected hidden layer that consists of 80 neurons with sigmoid activation functions, and we use L$_2$ regularisation with the weight of $l_2 = 0.001$. The output layer is also fully connected and has two outputs for learning the effective probability to be in two phases. Here the ReLU (rectified linear unit) activation functions are applied with $l_2 = 0.001$. The example of W-shape accuracy plot obtained during the learning stage for fixed $J$ and varying $W$ is shown in Fig.~\ref{fig:LbC}(a).

We apply LbC to the polariton lattice data and refine (and test) the boundaries of the phase diagram previously obtained from unsupervised learning. We concentrate on the parameter intervals where phase transition is potentially expected, and use LbC either to find the critical point of a transition, or merge phases if no W-shape dependence is observed. To train the NN we need use a large sample of polarisation patterns. This is achieved by taking a patch of parameters (working with the coarse-grained grid of $J$ and $W$) and generating multiple patterns by numerically solving Eq.~\eqref{eq.GP} for different initial conditions. 
The final phase diagram is shown in Fig.~\ref{fig:agglomerative}(b), which can be compared to the agglomerative clustering results in Fig.~\ref{fig:agglomerative}(a). At small $J$ and $W$ we reveal the region of linearly polarized condensates $\mathbf{s}_n = (-1,0,0)^\text{T}$ which we refer as the XY phase (labelled by \rom{1} and coloured in yellow). We note that this phase corresponds to the type of attractor shown in Fig.~\ref{fig_sketch}(b). As the pump gain increases, we approach three phases of different mixtures of ferromagnetic and antiferromagnetic ordering. We identify them in Fig.~\ref{fig:agglomerative}(b) as chequerboard AFM (\rom{2}), cluster AFM (\rom{3}), and the stripe phase (\rom{4}). At high $W$ and $J$ the system clearly enters the ferromagnetic phase (\rom{5}). At low $W$ the LbC scans however revealed only two phase boundaries where the W-shape emerges. We associate it to the diagonal stripe phase (\rom{6}) in the $-0.5<W<0.5$, $J>0.6$ region. At the same time, we did not identify distinct cluster with the conjectured glassy and wave phases sketched in Fig.~\ref{fig:tSNE}(b). We conclude that they likely correspond to the transition between phases (crossover). For instance, these are often observed in case of the finite-sized spin systems and manifest as domain walls and domain structures. We note that the performance of LbC procedure does not depend significantly on the structure of the NN, as long as it has a high expressivity. At the same time, the procedure is limited by the non-convex optimisation procedure (and possible local minimum trapping) as well as the fixed number of training samples that coarse-grains phase boundaries.

In the study we made the first steps towards mapping spin phases in polaritonic lattices. Exploiting a data-driven approach, we concentrated on clustering of polarisation patterns, and did not dive into the physics of identified phases. The next steps can include studying the identified diagonal stripe phase, highlighting the differences with respect to the horizontal/vertical stripe phase and other phases, and studies of cross-over to the FM phase. There are also potential ways to enhance the clustering. One route may be the analysis of data in the latent feature space obtained by variational autoencoders. Finally, learning by confusion approach can be further improved if deep NNs or more complex convolutional NNs are used.


\section*{Conclusions}

We have studied polarisation patterns that emerge as steady states in nonlinear polaritonic lattices. For different values of pump gain and lattice tunnelling rates, we see qualitatively distinct patterns that correspond to polariton phases with mixtures of ferromagnetic and antiferromagnetic bonding of chequerboard, stripe, diagonal and cluster types. Using data analysis and machine learning techniques we classified these patterns and identified their phase boundaries. First, a qualitative phase map is developed using the t-distributed stochastic neighbour embedding as a data visualisation tool. Next, unsupervised learning based on agglomerative clustering was used to sketch the phase diagram of polariton phases as a function tunnelling rate and pump gain. Finally, a neural network-based learning by confusion approach was used to mark and refine the boundaries between polariton phases. The work describes a path for studying phase transitions in nonlinear optical systems, and highlights the use of data-driven approaches in polaritonic systems.

\section*{Methods}

\subsection*{Numerical modelling}

To describe the dynamics of polaritonic lattices we solve Eq.~\eqref{eq.GP} for a square geometry with $8\times8$ sites with periodic boundary conditions. These are chosen to suppress the boundary effects, where close to thermodynamic limit physics can be studied. One can also consider soft-open boundary conditions, or damped boundary conditions, where dissipation grows as you get closer to the boundary. In the simulations, we vary two easily tunable experimental parameters $W_t$ and $J$ in the relevant range to generate a dataset of possible polarisation patterns accessible in experiment. The target nonresonant pump power $W$ can be readily tuned in time and the Josephson coupling strength $J$ can be tuned by changing the overlap between adjacent lattice sites at the lithography stage (micropillars), or by tuning the lattice potential optically.

\subsection*{Visualisation}

We analyse the dataset using the open source Python library {\sffamily{}sklearn}. We perform t-SNE with adjustable hyperparameters being the perplexity and the learning rate. Perplexity corresponds to the averaged number of accounted nearest neighbours (data points) which affect the learning process, and generally sets the statistical certainty in separating two points. Learning rate is responsible for the rate at which we update the positions, determining the step size in minimisation of loss function. The hyperparameters can be tuned to balance the capture of local and global details in the dataset. A good choice of hyperparameters can be additionally tested by confirming the effective clustering of known polaritonic phases with ferromagnetic and antiferromagnetic patterns (for instance, by labelling known configurations and checking their positions on the t-SNE diagram).

\subsection*{Clustering}

We use $K$-means and agglomerative clustering approaches~\cite{Mehta2019}. Both methods generally search for the mean values for $K$ clusters, and adjust those means such that the Euclidean distance between the means and data points are minimised. $K$-means clustering requires defining the number of clusters $K$ in advance. In contrast, the agglomerative clustering belongs to hierarchical methods. At first, all data points are assigned to distinct clusters (labelled from $1$ up to the cardinality of the data point $\bm{v}$). Next, using the pre-defined distance metric for two data points $\bm{v}$ and $\bm{w}$ from different cluster, the difference between clusters is evaluated. The cluster with a difference being below the threshold value are merged iteratively. The distance corresponds to four distinct types: complete, single, average, and Ward's. The complete distance type relies on the maximum distance between two data points in different clusters. The single distance type uses the minimum distance between two points from different clusters. The average distance type relies on the average distance between all of points from two clusters that are compared. The ward distance type relies on the sum of squared distances to the centre of the cluster. The popular distance metrics are: 1) Euclidean distance $d(\bm{v}, \bm{w}) := \lVert \bm{v} - \bm{w} \rVert$ as L$_2$ norm of the difference of two vectors; 2) cosine distance $d(\bm{v}, \bm{w}) := \bm{v} \cdot \bm{w}/( {\lVert \bm{v}\rVert ~ \lVert \bm{w}\rVert})$; 3) Manhattan L$_1$ distance $d(\bm{v}, \bm{w}) := \sum_{i}{|v_{i} - w_{i}|}$; among others.

\begin{acknowledgments}
O.\,K. thanks Oleksandr Balabanov, Bart Olsthoorn, and Stanislav Borysov for useful suggestions. 
I.\,A.\,S. and I.\,I. acknowledge the support from Russian Foundation for Basic Research (RFBR), in framework of the joint RFBR-DFG project No. 21-52-12038. O.\,K. acknowledges the support from UK EPSRC New Investigator Award under the Agreement No. EP/ V00171X/1. H.\,S. acknowledges the support of the UK’s Engineering and Physical Sciences Research Council (grant EP/M025330/1 on Hybrid Polaritonics), and European Union’s Horizon 2020 program, through a FET Open research and innovation action under the grant agreement No. 899141 (PoLLoC), and the Icelandic Research Fund (Rannis), grant No. 217631-051. V.\,K.\,K and I.\,A.\,S. acknowledge support from Icelandic Research Fund (Rannis), grant No. 163082-051.
\end{acknowledgments}

\begin{center}\textbf{AUTHOR CONTRIBUTIONS}\end{center}\vspace{-2mm}
D.\,Z. and H.\,S. performed the calculations for ML/clustering and polarisation dynamics, respectively. O.\,K. and H.\,S. written the manuscript, with contributions from D.\,Z. and I.\,A.\,S. V.\,K.\,K. and I.\,I. provided crucial suggestions on polarisation pattern clustering and dataset preprocessing. All authors contributed to discussing results, with  O.\,K., I.\,A.\,S., and V.\,U. managed the project. \vspace{1mm}

\begin{center}\textbf{COMPETING INTERESTS}\end{center}\vspace{-2mm}
The authors declare no competing interests.\vspace{1mm}

\begin{center}\textbf{DATA AVAILABILITY}\end{center}\vspace{-2mm}
The data that support the findings of this study are available from the corresponding author upon reasonable request.\vspace{1mm}

\begin{center}\textbf{CODE AVAILABILITY}\end{center}\vspace{-2mm}
The code for the analysis is available from the corresponding author upon reasonable request.\vspace{1mm}


%


\begin{thebibliography}{88}%
\makeatletter
\providecommand \@ifxundefined [1]{%
 \@ifx{#1\undefined}
}%
\providecommand \@ifnum [1]{%
 \ifnum #1\expandafter \@firstoftwo
 \else \expandafter \@secondoftwo
 \fi
}%
\providecommand \@ifx [1]{%
 \ifx #1\expandafter \@firstoftwo
 \else \expandafter \@secondoftwo
 \fi
}%
\providecommand \natexlab [1]{#1}%
\providecommand \enquote  [1]{``#1''}%
\providecommand \bibnamefont  [1]{#1}%
\providecommand \bibfnamefont [1]{#1}%
\providecommand \citenamefont [1]{#1}%
\providecommand \href@noop [0]{\@secondoftwo}%
\providecommand \href [0]{\begingroup \@sanitize@url \@href}%
\providecommand \@href[1]{\@@startlink{#1}\@@href}%
\providecommand \@@href[1]{\endgroup#1\@@endlink}%
\providecommand \@sanitize@url [0]{\catcode `\\12\catcode `\$12\catcode
  `\&12\catcode `\#12\catcode `\^12\catcode `\_12\catcode `\%12\relax}%
\providecommand \@@startlink[1]{}%
\providecommand \@@endlink[0]{}%
\providecommand \url  [0]{\begingroup\@sanitize@url \@url }%
\providecommand \@url [1]{\endgroup\@href {#1}{\urlprefix }}%
\providecommand \urlprefix  [0]{URL }%
\providecommand \Eprint [0]{\href }%
\providecommand \doibase [0]{https://doi.org/}%
\providecommand \selectlanguage [0]{\@gobble}%
\providecommand \bibinfo  [0]{\@secondoftwo}%
\providecommand \bibfield  [0]{\@secondoftwo}%
\providecommand \translation [1]{[#1]}%
\providecommand \BibitemOpen [0]{}%
\providecommand \bibitemStop [0]{}%
\providecommand \bibitemNoStop [0]{.\EOS\space}%
\providecommand \EOS [0]{\spacefactor3000\relax}%
\providecommand \BibitemShut  [1]{\csname bibitem#1\endcsname}%
\let\auto@bib@innerbib\@empty
\bibitem [{\citenamefont {Carleo}\ \emph {et~al.}(2019)\citenamefont {Carleo},
  \citenamefont {Cirac}, \citenamefont {Cranmer}, \citenamefont {Daudet},
  \citenamefont {Schuld}, \citenamefont {Tishby}, \citenamefont
  {Vogt-Maranto},\ and\ \citenamefont {Zdeborov\'a}}]{Carleo2019}%
  \BibitemOpen
  \bibfield  {author} {\bibinfo {author} {\bibfnamefont {G.}~\bibnamefont
  {Carleo}}, \bibinfo {author} {\bibfnamefont {I.}~\bibnamefont {Cirac}},
  \bibinfo {author} {\bibfnamefont {K.}~\bibnamefont {Cranmer}}, \bibinfo
  {author} {\bibfnamefont {L.}~\bibnamefont {Daudet}}, \bibinfo {author}
  {\bibfnamefont {M.}~\bibnamefont {Schuld}}, \bibinfo {author} {\bibfnamefont
  {N.}~\bibnamefont {Tishby}}, \bibinfo {author} {\bibfnamefont
  {L.}~\bibnamefont {Vogt-Maranto}},\ and\ \bibinfo {author} {\bibfnamefont
  {L.}~\bibnamefont {Zdeborov\'a}},\ }\bibfield  {title} {\bibinfo {title}
  {Machine learning and the physical sciences},\ }\href
  {https://doi.org/10.1103/RevModPhys.91.045002} {\bibfield  {journal}
  {\bibinfo  {journal} {Rev. Mod. Phys.}\ }\textbf {\bibinfo {volume} {91}},\
  \bibinfo {pages} {045002} (\bibinfo {year} {2019})}\BibitemShut {NoStop}%
\bibitem [{\citenamefont {Mehta}\ \emph {et~al.}(2019)\citenamefont {Mehta},
  \citenamefont {Bukov}, \citenamefont {Wang}, \citenamefont {Day},
  \citenamefont {Richardson}, \citenamefont {Fisher},\ and\ \citenamefont
  {Schwab}}]{Mehta2019}%
  \BibitemOpen
  \bibfield  {author} {\bibinfo {author} {\bibfnamefont {P.}~\bibnamefont
  {Mehta}}, \bibinfo {author} {\bibfnamefont {M.}~\bibnamefont {Bukov}},
  \bibinfo {author} {\bibfnamefont {C.-H.}\ \bibnamefont {Wang}}, \bibinfo
  {author} {\bibfnamefont {A.~G.}\ \bibnamefont {Day}}, \bibinfo {author}
  {\bibfnamefont {C.}~\bibnamefont {Richardson}}, \bibinfo {author}
  {\bibfnamefont {C.~K.}\ \bibnamefont {Fisher}},\ and\ \bibinfo {author}
  {\bibfnamefont {D.~J.}\ \bibnamefont {Schwab}},\ }\bibfield  {title}
  {\bibinfo {title} {A high-bias, low-variance introduction to machine learning
  for physicists},\ }\href
  {https://doi.org/https://doi.org/10.1016/j.physrep.2019.03.001} {\bibfield
  {journal} {\bibinfo  {journal} {Physics Reports}\ }\textbf {\bibinfo {volume}
  {810}},\ \bibinfo {pages} {1} (\bibinfo {year} {2019})},\ \bibinfo {note} {a
  high-bias, low-variance introduction to Machine Learning for
  physicists}\BibitemShut {NoStop}%
\bibitem [{\citenamefont {{Goodfellow}}\ \emph {et~al.}(2014)\citenamefont
  {{Goodfellow}}, \citenamefont {{Pouget-Abadie}}, \citenamefont {{Mirza}},
  \citenamefont {{Xu}}, \citenamefont {{Warde-Farley}}, \citenamefont
  {{Ozair}}, \citenamefont {{Courville}},\ and\ \citenamefont
  {{Bengio}}}]{Goodfellow2014}%
  \BibitemOpen
  \bibfield  {author} {\bibinfo {author} {\bibfnamefont {I.~J.}\ \bibnamefont
  {{Goodfellow}}}, \bibinfo {author} {\bibfnamefont {J.}~\bibnamefont
  {{Pouget-Abadie}}}, \bibinfo {author} {\bibfnamefont {M.}~\bibnamefont
  {{Mirza}}}, \bibinfo {author} {\bibfnamefont {B.}~\bibnamefont {{Xu}}},
  \bibinfo {author} {\bibfnamefont {D.}~\bibnamefont {{Warde-Farley}}},
  \bibinfo {author} {\bibfnamefont {S.}~\bibnamefont {{Ozair}}}, \bibinfo
  {author} {\bibfnamefont {A.}~\bibnamefont {{Courville}}},\ and\ \bibinfo
  {author} {\bibfnamefont {Y.}~\bibnamefont {{Bengio}}},\ }\bibfield  {title}
  {\bibinfo {title} {{Generative Adversarial Networks}},\ }\href@noop {}
  {\bibfield  {journal} {\bibinfo  {journal} {arXiv e-prints}\ ,\ \bibinfo
  {eid} {arXiv:1406.2661}} (\bibinfo {year} {2014})},\ \Eprint
  {https://arxiv.org/abs/1406.2661} {arXiv:1406.2661 [stat.ML]} \BibitemShut
  {NoStop}%
\bibitem [{\citenamefont {Gomez-Uribe}\ and\ \citenamefont
  {Hunt}(2016)}]{Netflix2015}%
  \BibitemOpen
  \bibfield  {author} {\bibinfo {author} {\bibfnamefont {C.~A.}\ \bibnamefont
  {Gomez-Uribe}}\ and\ \bibinfo {author} {\bibfnamefont {N.}~\bibnamefont
  {Hunt}},\ }\bibfield  {title} {\bibinfo {title} {The netflix recommender
  system: Algorithms, business value, and innovation},\ }\bibfield  {journal}
  {\bibinfo  {journal} {ACM Trans. Manage. Inf. Syst.}\ }\textbf {\bibinfo
  {volume} {6}},\ \href {https://doi.org/10.1145/2843948} {10.1145/2843948}
  (\bibinfo {year} {2016})\BibitemShut {NoStop}%
\bibitem [{\citenamefont {{Vaswani}}\ \emph {et~al.}(2017)\citenamefont
  {{Vaswani}}, \citenamefont {{Shazeer}}, \citenamefont {{Parmar}},
  \citenamefont {{Uszkoreit}}, \citenamefont {{Jones}}, \citenamefont
  {{Gomez}}, \citenamefont {{Kaiser}},\ and\ \citenamefont
  {{Polosukhin}}}]{NLP2017}%
  \BibitemOpen
  \bibfield  {author} {\bibinfo {author} {\bibfnamefont {A.}~\bibnamefont
  {{Vaswani}}}, \bibinfo {author} {\bibfnamefont {N.}~\bibnamefont
  {{Shazeer}}}, \bibinfo {author} {\bibfnamefont {N.}~\bibnamefont {{Parmar}}},
  \bibinfo {author} {\bibfnamefont {J.}~\bibnamefont {{Uszkoreit}}}, \bibinfo
  {author} {\bibfnamefont {L.}~\bibnamefont {{Jones}}}, \bibinfo {author}
  {\bibfnamefont {A.~N.}\ \bibnamefont {{Gomez}}}, \bibinfo {author}
  {\bibfnamefont {L.}~\bibnamefont {{Kaiser}}},\ and\ \bibinfo {author}
  {\bibfnamefont {I.}~\bibnamefont {{Polosukhin}}},\ }\bibfield  {title}
  {\bibinfo {title} {{Attention Is All You Need}},\ }\href@noop {} {\bibfield
  {journal} {\bibinfo  {journal} {arXiv e-prints}\ ,\ \bibinfo {eid}
  {arXiv:1706.03762}} (\bibinfo {year} {2017})},\ \Eprint
  {https://arxiv.org/abs/1706.03762} {arXiv:1706.03762 [cs.CL]} \BibitemShut
  {NoStop}%
\bibitem [{\citenamefont {Liu}\ \emph {et~al.}(2019)\citenamefont {Liu},
  \citenamefont {Faes}, \citenamefont {Kale}, \citenamefont {Wagner},
  \citenamefont {Fu}, \citenamefont {Bruynseels}, \citenamefont {Mahendiran},
  \citenamefont {Moraes}, \citenamefont {Shamdas}, \citenamefont {Kern},
  \citenamefont {Ledsam}, \citenamefont {Schmid}, \citenamefont {Balaskas},
  \citenamefont {Topol}, \citenamefont {Bachmann}, \citenamefont {Keane},\ and\
  \citenamefont {Denniston}}]{Lancet2019}%
  \BibitemOpen
  \bibfield  {author} {\bibinfo {author} {\bibfnamefont {X.}~\bibnamefont
  {Liu}}, \bibinfo {author} {\bibfnamefont {L.}~\bibnamefont {Faes}}, \bibinfo
  {author} {\bibfnamefont {A.~U.}\ \bibnamefont {Kale}}, \bibinfo {author}
  {\bibfnamefont {S.~K.}\ \bibnamefont {Wagner}}, \bibinfo {author}
  {\bibfnamefont {D.~J.}\ \bibnamefont {Fu}}, \bibinfo {author} {\bibfnamefont
  {A.}~\bibnamefont {Bruynseels}}, \bibinfo {author} {\bibfnamefont
  {T.}~\bibnamefont {Mahendiran}}, \bibinfo {author} {\bibfnamefont
  {G.}~\bibnamefont {Moraes}}, \bibinfo {author} {\bibfnamefont
  {M.}~\bibnamefont {Shamdas}}, \bibinfo {author} {\bibfnamefont
  {C.}~\bibnamefont {Kern}}, \bibinfo {author} {\bibfnamefont {J.~R.}\
  \bibnamefont {Ledsam}}, \bibinfo {author} {\bibfnamefont {M.~K.}\
  \bibnamefont {Schmid}}, \bibinfo {author} {\bibfnamefont {K.}~\bibnamefont
  {Balaskas}}, \bibinfo {author} {\bibfnamefont {E.~J.}\ \bibnamefont {Topol}},
  \bibinfo {author} {\bibfnamefont {L.~M.}\ \bibnamefont {Bachmann}}, \bibinfo
  {author} {\bibfnamefont {P.~A.}\ \bibnamefont {Keane}},\ and\ \bibinfo
  {author} {\bibfnamefont {A.~K.}\ \bibnamefont {Denniston}},\ }\bibfield
  {title} {\bibinfo {title} {A comparison of deep learning performance against
  health-care professionals in detecting diseases from medical imaging: a
  systematic review and meta-analysis},\ }\href
  {https://doi.org/10.1016/S2589-7500(19)30123-2} {\bibfield  {journal}
  {\bibinfo  {journal} {The Lancet Digital Health}\ }\textbf {\bibinfo {volume}
  {1}},\ \bibinfo {pages} {e271} (\bibinfo {year} {2019})}\BibitemShut
  {NoStop}%
\bibitem [{\citenamefont {Benedetti}\ \emph {et~al.}(2019)\citenamefont
  {Benedetti}, \citenamefont {Lloyd}, \citenamefont {Sack},\ and\ \citenamefont
  {Fiorentini}}]{Benedetti2019}%
  \BibitemOpen
  \bibfield  {author} {\bibinfo {author} {\bibfnamefont {M.}~\bibnamefont
  {Benedetti}}, \bibinfo {author} {\bibfnamefont {E.}~\bibnamefont {Lloyd}},
  \bibinfo {author} {\bibfnamefont {S.}~\bibnamefont {Sack}},\ and\ \bibinfo
  {author} {\bibfnamefont {M.}~\bibnamefont {Fiorentini}},\ }\bibfield  {title}
  {\bibinfo {title} {Parameterized quantum circuits as machine learning
  models},\ }\href {https://doi.org/10.1088/2058-9565/ab4eb5} {\bibfield
  {journal} {\bibinfo  {journal} {Quantum Science and Technology}\ }\textbf
  {\bibinfo {volume} {4}},\ \bibinfo {pages} {043001} (\bibinfo {year}
  {2019})}\BibitemShut {NoStop}%
\bibitem [{\citenamefont {Uvarov}\ \emph {et~al.}(2020)\citenamefont {Uvarov},
  \citenamefont {Kardashin},\ and\ \citenamefont {Biamonte}}]{Uvarov2020}%
  \BibitemOpen
  \bibfield  {author} {\bibinfo {author} {\bibfnamefont {A.~V.}\ \bibnamefont
  {Uvarov}}, \bibinfo {author} {\bibfnamefont {A.~S.}\ \bibnamefont
  {Kardashin}},\ and\ \bibinfo {author} {\bibfnamefont {J.~D.}\ \bibnamefont
  {Biamonte}},\ }\bibfield  {title} {\bibinfo {title} {Machine learning phase
  transitions with a quantum processor},\ }\href
  {https://doi.org/10.1103/PhysRevA.102.012415} {\bibfield  {journal} {\bibinfo
   {journal} {Phys. Rev. A}\ }\textbf {\bibinfo {volume} {102}},\ \bibinfo
  {pages} {012415} (\bibinfo {year} {2020})}\BibitemShut {NoStop}%
\bibitem [{\citenamefont {Carrasquilla}\ and\ \citenamefont
  {Melko}(2017)}]{CarrasquillaMelko}%
  \BibitemOpen
  \bibfield  {author} {\bibinfo {author} {\bibfnamefont {J.}~\bibnamefont
  {Carrasquilla}}\ and\ \bibinfo {author} {\bibfnamefont {R.~G.}\ \bibnamefont
  {Melko}},\ }\bibfield  {title} {\bibinfo {title} {Machine learning phases of
  matter},\ }\href {https://doi.org/10.1038/nphys4035} {\bibfield  {journal}
  {\bibinfo  {journal} {Nature Physics}\ }\textbf {\bibinfo {volume} {13}},\
  \bibinfo {pages} {431} (\bibinfo {year} {2017})}\BibitemShut {NoStop}%
\bibitem [{\citenamefont {Wang}(2016)}]{Wang2016}%
  \BibitemOpen
  \bibfield  {author} {\bibinfo {author} {\bibfnamefont {L.}~\bibnamefont
  {Wang}},\ }\bibfield  {title} {\bibinfo {title} {Discovering phase
  transitions with unsupervised learning},\ }\href
  {https://doi.org/10.1103/PhysRevB.94.195105} {\bibfield  {journal} {\bibinfo
  {journal} {Phys. Rev. B}\ }\textbf {\bibinfo {volume} {94}},\ \bibinfo
  {pages} {195105} (\bibinfo {year} {2016})}\BibitemShut {NoStop}%
\bibitem [{\citenamefont {Torlai}\ and\ \citenamefont
  {Melko}(2016)}]{Torlai2016}%
  \BibitemOpen
  \bibfield  {author} {\bibinfo {author} {\bibfnamefont {G.}~\bibnamefont
  {Torlai}}\ and\ \bibinfo {author} {\bibfnamefont {R.~G.}\ \bibnamefont
  {Melko}},\ }\bibfield  {title} {\bibinfo {title} {Learning thermodynamics
  with boltzmann machines},\ }\href
  {https://doi.org/10.1103/PhysRevB.94.165134} {\bibfield  {journal} {\bibinfo
  {journal} {Phys. Rev. B}\ }\textbf {\bibinfo {volume} {94}},\ \bibinfo
  {pages} {165134} (\bibinfo {year} {2016})}\BibitemShut {NoStop}%
\bibitem [{\citenamefont {Hu}\ \emph {et~al.}(2017)\citenamefont {Hu},
  \citenamefont {Singh},\ and\ \citenamefont {Scalettar}}]{WHu2017}%
  \BibitemOpen
  \bibfield  {author} {\bibinfo {author} {\bibfnamefont {W.}~\bibnamefont
  {Hu}}, \bibinfo {author} {\bibfnamefont {R.~R.~P.}\ \bibnamefont {Singh}},\
  and\ \bibinfo {author} {\bibfnamefont {R.~T.}\ \bibnamefont {Scalettar}},\
  }\bibfield  {title} {\bibinfo {title} {Discovering phases, phase transitions,
  and crossovers through unsupervised machine learning: A critical
  examination},\ }\href {https://doi.org/10.1103/PhysRevE.95.062122} {\bibfield
   {journal} {\bibinfo  {journal} {Phys. Rev. E}\ }\textbf {\bibinfo {volume}
  {95}},\ \bibinfo {pages} {062122} (\bibinfo {year} {2017})}\BibitemShut
  {NoStop}%
\bibitem [{\citenamefont {van Nieuwenburg}\ \emph {et~al.}(2017)\citenamefont
  {van Nieuwenburg}, \citenamefont {Liu},\ and\ \citenamefont
  {Huber}}]{Nieuwenburg2017}%
  \BibitemOpen
  \bibfield  {author} {\bibinfo {author} {\bibfnamefont {E.~P.~L.}\
  \bibnamefont {van Nieuwenburg}}, \bibinfo {author} {\bibfnamefont {Y.-H.}\
  \bibnamefont {Liu}},\ and\ \bibinfo {author} {\bibfnamefont {S.~D.}\
  \bibnamefont {Huber}},\ }\bibfield  {title} {\bibinfo {title} {Learning phase
  transitions by confusion},\ }\href {https://doi.org/10.1038/nphys4037}
  {\bibfield  {journal} {\bibinfo  {journal} {Nature Physics}\ }\textbf
  {\bibinfo {volume} {13}},\ \bibinfo {pages} {435} (\bibinfo {year}
  {2017})}\BibitemShut {NoStop}%
\bibitem [{\citenamefont {Shirinyan}\ \emph {et~al.}(2019)\citenamefont
  {Shirinyan}, \citenamefont {Kozin}, \citenamefont {Hellsvik}, \citenamefont
  {Pereiro}, \citenamefont {Eriksson},\ and\ \citenamefont
  {Yudin}}]{KozinVK_ML_2019}%
  \BibitemOpen
  \bibfield  {author} {\bibinfo {author} {\bibfnamefont {A.~A.}\ \bibnamefont
  {Shirinyan}}, \bibinfo {author} {\bibfnamefont {V.~K.}\ \bibnamefont
  {Kozin}}, \bibinfo {author} {\bibfnamefont {J.}~\bibnamefont {Hellsvik}},
  \bibinfo {author} {\bibfnamefont {M.}~\bibnamefont {Pereiro}}, \bibinfo
  {author} {\bibfnamefont {O.}~\bibnamefont {Eriksson}},\ and\ \bibinfo
  {author} {\bibfnamefont {D.}~\bibnamefont {Yudin}},\ }\bibfield  {title}
  {\bibinfo {title} {Self-organizing maps as a method for detecting phase
  transitions and phase identification},\ }\href
  {https://doi.org/10.1103/PhysRevB.99.041108} {\bibfield  {journal} {\bibinfo
  {journal} {Phys. Rev. B}\ }\textbf {\bibinfo {volume} {99}},\ \bibinfo
  {pages} {041108} (\bibinfo {year} {2019})}\BibitemShut {NoStop}%
\bibitem [{\citenamefont {{Corte}}\ \emph {et~al.}(2020)\citenamefont
  {{Corte}}, \citenamefont {{Acevedo}}, \citenamefont {{Arlego}},\ and\
  \citenamefont {{Lamas}}}]{Corte2020}%
  \BibitemOpen
  \bibfield  {author} {\bibinfo {author} {\bibfnamefont {I.}~\bibnamefont
  {{Corte}}}, \bibinfo {author} {\bibfnamefont {S.}~\bibnamefont {{Acevedo}}},
  \bibinfo {author} {\bibfnamefont {M.}~\bibnamefont {{Arlego}}},\ and\
  \bibinfo {author} {\bibfnamefont {C.~A.}\ \bibnamefont {{Lamas}}},\
  }\bibfield  {title} {\bibinfo {title} {{Transfer and confusion deep learning
  in frustrated spin systems}},\ }\href@noop {} {\bibfield  {journal} {\bibinfo
   {journal} {arXiv e-prints}\ ,\ \bibinfo {eid} {arXiv:2009.00661}} (\bibinfo
  {year} {2020})},\ \Eprint {https://arxiv.org/abs/2009.00661}
  {arXiv:2009.00661 [physics.comp-ph]} \BibitemShut {NoStop}%
\bibitem [{\citenamefont {Deng}\ \emph {et~al.}(2017)\citenamefont {Deng},
  \citenamefont {Li},\ and\ \citenamefont {Das~Sarma}}]{Deng2017}%
  \BibitemOpen
  \bibfield  {author} {\bibinfo {author} {\bibfnamefont {D.-L.}\ \bibnamefont
  {Deng}}, \bibinfo {author} {\bibfnamefont {X.}~\bibnamefont {Li}},\ and\
  \bibinfo {author} {\bibfnamefont {S.}~\bibnamefont {Das~Sarma}},\ }\bibfield
  {title} {\bibinfo {title} {Machine learning topological states},\ }\href
  {https://doi.org/10.1103/PhysRevB.96.195145} {\bibfield  {journal} {\bibinfo
  {journal} {Phys. Rev. B}\ }\textbf {\bibinfo {volume} {96}},\ \bibinfo
  {pages} {195145} (\bibinfo {year} {2017})}\BibitemShut {NoStop}%
\bibitem [{\citenamefont {Rodriguez-Nieva}\ and\ \citenamefont
  {Scheurer}(2019)}]{Rodriguez2019}%
  \BibitemOpen
  \bibfield  {author} {\bibinfo {author} {\bibfnamefont {J.~F.}\ \bibnamefont
  {Rodriguez-Nieva}}\ and\ \bibinfo {author} {\bibfnamefont {M.~S.}\
  \bibnamefont {Scheurer}},\ }\bibfield  {title} {\bibinfo {title} {Identifying
  topological order through unsupervised machine learning},\ }\href
  {https://doi.org/10.1038/s41567-019-0512-x} {\bibfield  {journal} {\bibinfo
  {journal} {Nature Physics}\ }\textbf {\bibinfo {volume} {15}},\ \bibinfo
  {pages} {790} (\bibinfo {year} {2019})}\BibitemShut {NoStop}%
\bibitem [{\citenamefont {Zhang}\ \emph {et~al.}(2019)\citenamefont {Zhang},
  \citenamefont {Liu},\ and\ \citenamefont {Wei}}]{TZWei2019}%
  \BibitemOpen
  \bibfield  {author} {\bibinfo {author} {\bibfnamefont {W.}~\bibnamefont
  {Zhang}}, \bibinfo {author} {\bibfnamefont {J.}~\bibnamefont {Liu}},\ and\
  \bibinfo {author} {\bibfnamefont {T.-C.}\ \bibnamefont {Wei}},\ }\bibfield
  {title} {\bibinfo {title} {Machine learning of phase transitions in the
  percolation and $xy$ models},\ }\href
  {https://doi.org/10.1103/PhysRevE.99.032142} {\bibfield  {journal} {\bibinfo
  {journal} {Phys. Rev. E}\ }\textbf {\bibinfo {volume} {99}},\ \bibinfo
  {pages} {032142} (\bibinfo {year} {2019})}\BibitemShut {NoStop}%
\bibitem [{\citenamefont {Canabarro}\ \emph {et~al.}(2019)\citenamefont
  {Canabarro}, \citenamefont {Fanchini}, \citenamefont {Malvezzi},
  \citenamefont {Pereira},\ and\ \citenamefont {Chaves}}]{Canabarro2019}%
  \BibitemOpen
  \bibfield  {author} {\bibinfo {author} {\bibfnamefont {A.}~\bibnamefont
  {Canabarro}}, \bibinfo {author} {\bibfnamefont {F.~F.}\ \bibnamefont
  {Fanchini}}, \bibinfo {author} {\bibfnamefont {A.~L.}\ \bibnamefont
  {Malvezzi}}, \bibinfo {author} {\bibfnamefont {R.}~\bibnamefont {Pereira}},\
  and\ \bibinfo {author} {\bibfnamefont {R.}~\bibnamefont {Chaves}},\
  }\bibfield  {title} {\bibinfo {title} {Unveiling phase transitions with
  machine learning},\ }\href {https://doi.org/10.1103/PhysRevB.100.045129}
  {\bibfield  {journal} {\bibinfo  {journal} {Phys. Rev. B}\ }\textbf {\bibinfo
  {volume} {100}},\ \bibinfo {pages} {045129} (\bibinfo {year}
  {2019})}\BibitemShut {NoStop}%
\bibitem [{\citenamefont {Balabanov}\ and\ \citenamefont
  {Granath}(2020)}]{Balabanov2020}%
  \BibitemOpen
  \bibfield  {author} {\bibinfo {author} {\bibfnamefont {O.}~\bibnamefont
  {Balabanov}}\ and\ \bibinfo {author} {\bibfnamefont {M.}~\bibnamefont
  {Granath}},\ }\bibfield  {title} {\bibinfo {title} {Unsupervised learning
  using topological data augmentation},\ }\href
  {https://doi.org/10.1103/PhysRevResearch.2.013354} {\bibfield  {journal}
  {\bibinfo  {journal} {Phys. Rev. Research}\ }\textbf {\bibinfo {volume}
  {2}},\ \bibinfo {pages} {013354} (\bibinfo {year} {2020})}\BibitemShut
  {NoStop}%
\bibitem [{\citenamefont {Scheurer}\ and\ \citenamefont
  {Slager}(2020)}]{Scheurer2020}%
  \BibitemOpen
  \bibfield  {author} {\bibinfo {author} {\bibfnamefont {M.~S.}\ \bibnamefont
  {Scheurer}}\ and\ \bibinfo {author} {\bibfnamefont {R.-J.}\ \bibnamefont
  {Slager}},\ }\bibfield  {title} {\bibinfo {title} {Unsupervised machine
  learning and band topology},\ }\href
  {https://doi.org/10.1103/PhysRevLett.124.226401} {\bibfield  {journal}
  {\bibinfo  {journal} {Phys. Rev. Lett.}\ }\textbf {\bibinfo {volume} {124}},\
  \bibinfo {pages} {226401} (\bibinfo {year} {2020})}\BibitemShut {NoStop}%
\bibitem [{\citenamefont {Rodrigues}\ \emph {et~al.}(2021)\citenamefont
  {Rodrigues}, \citenamefont {Dhar}, \citenamefont {Walker}, \citenamefont
  {Smith}, \citenamefont {Oulton}, \citenamefont {Mintert},\ and\ \citenamefont
  {Nyman}}]{Rodrigues_PRL2021}%
  \BibitemOpen
  \bibfield  {author} {\bibinfo {author} {\bibfnamefont {J.~D.}\ \bibnamefont
  {Rodrigues}}, \bibinfo {author} {\bibfnamefont {H.~S.}\ \bibnamefont {Dhar}},
  \bibinfo {author} {\bibfnamefont {B.~T.}\ \bibnamefont {Walker}}, \bibinfo
  {author} {\bibfnamefont {J.~M.}\ \bibnamefont {Smith}}, \bibinfo {author}
  {\bibfnamefont {R.~F.}\ \bibnamefont {Oulton}}, \bibinfo {author}
  {\bibfnamefont {F.}~\bibnamefont {Mintert}},\ and\ \bibinfo {author}
  {\bibfnamefont {R.~A.}\ \bibnamefont {Nyman}},\ }\bibfield  {title} {\bibinfo
  {title} {Learning the fuzzy phases of small photonic condensates},\ }\href
  {https://doi.org/10.1103/PhysRevLett.126.150602} {\bibfield  {journal}
  {\bibinfo  {journal} {Phys. Rev. Lett.}\ }\textbf {\bibinfo {volume} {126}},\
  \bibinfo {pages} {150602} (\bibinfo {year} {2021})}\BibitemShut {NoStop}%
\bibitem [{\citenamefont {Ohtsuki}\ and\ \citenamefont
  {Ohtsuki}(2016)}]{Ohtsuki2016}%
  \BibitemOpen
  \bibfield  {author} {\bibinfo {author} {\bibfnamefont {T.}~\bibnamefont
  {Ohtsuki}}\ and\ \bibinfo {author} {\bibfnamefont {T.}~\bibnamefont
  {Ohtsuki}},\ }\bibfield  {title} {\bibinfo {title} {Deep learning the quantum
  phase transitions in random two-dimensional electron systems},\ }\href
  {https://doi.org/10.7566/JPSJ.85.123706} {\bibfield  {journal} {\bibinfo
  {journal} {Journal of the Physical Society of Japan}\ }\textbf {\bibinfo
  {volume} {85}},\ \bibinfo {pages} {123706} (\bibinfo {year}
  {2016})}\BibitemShut {NoStop}%
\bibitem [{\citenamefont {Ohtsuki}\ and\ \citenamefont
  {Ohtsuki}(2017)}]{Ohtsuki2017}%
  \BibitemOpen
  \bibfield  {author} {\bibinfo {author} {\bibfnamefont {T.}~\bibnamefont
  {Ohtsuki}}\ and\ \bibinfo {author} {\bibfnamefont {T.}~\bibnamefont
  {Ohtsuki}},\ }\bibfield  {title} {\bibinfo {title} {Deep learning the quantum
  phase transitions in random electron systems: Applications to three
  dimensions},\ }\href {https://doi.org/10.7566/JPSJ.86.044708} {\bibfield
  {journal} {\bibinfo  {journal} {Journal of the Physical Society of Japan}\
  }\textbf {\bibinfo {volume} {86}},\ \bibinfo {pages} {044708} (\bibinfo
  {year} {2017})}\BibitemShut {NoStop}%
\bibitem [{\citenamefont {Ch'ng}\ \emph {et~al.}(2017)\citenamefont {Ch'ng},
  \citenamefont {Carrasquilla}, \citenamefont {Melko},\ and\ \citenamefont
  {Khatami}}]{Chng2017}%
  \BibitemOpen
  \bibfield  {author} {\bibinfo {author} {\bibfnamefont {K.}~\bibnamefont
  {Ch'ng}}, \bibinfo {author} {\bibfnamefont {J.}~\bibnamefont {Carrasquilla}},
  \bibinfo {author} {\bibfnamefont {R.~G.}\ \bibnamefont {Melko}},\ and\
  \bibinfo {author} {\bibfnamefont {E.}~\bibnamefont {Khatami}},\ }\bibfield
  {title} {\bibinfo {title} {Machine learning phases of strongly correlated
  fermions},\ }\href {https://doi.org/10.1103/PhysRevX.7.031038} {\bibfield
  {journal} {\bibinfo  {journal} {Phys. Rev. X}\ }\textbf {\bibinfo {volume}
  {7}},\ \bibinfo {pages} {031038} (\bibinfo {year} {2017})}\BibitemShut
  {NoStop}%
\bibitem [{\citenamefont {Borysov}\ \emph {et~al.}(2018)\citenamefont
  {Borysov}, \citenamefont {Olsthoorn}, \citenamefont {Gedik}, \citenamefont
  {Geilhufe},\ and\ \citenamefont {Balatsky}}]{Borysov2018}%
  \BibitemOpen
  \bibfield  {author} {\bibinfo {author} {\bibfnamefont {S.~S.}\ \bibnamefont
  {Borysov}}, \bibinfo {author} {\bibfnamefont {B.}~\bibnamefont {Olsthoorn}},
  \bibinfo {author} {\bibfnamefont {M.~B.}\ \bibnamefont {Gedik}}, \bibinfo
  {author} {\bibfnamefont {R.~M.}\ \bibnamefont {Geilhufe}},\ and\ \bibinfo
  {author} {\bibfnamefont {A.~V.}\ \bibnamefont {Balatsky}},\ }\bibfield
  {title} {\bibinfo {title} {Online search tool for graphical patterns in
  electronic band structures},\ }\href
  {https://doi.org/10.1038/s41524-018-0104-9} {\bibfield  {journal} {\bibinfo
  {journal} {npj Computational Materials}\ }\textbf {\bibinfo {volume} {4}},\
  \bibinfo {pages} {46} (\bibinfo {year} {2018})}\BibitemShut {NoStop}%
\bibitem [{\citenamefont {Olsthoorn}\ \emph {et~al.}(2019)\citenamefont
  {Olsthoorn}, \citenamefont {Geilhufe}, \citenamefont {Borysov},\ and\
  \citenamefont {Balatsky}}]{Olsthoorn2019}%
  \BibitemOpen
  \bibfield  {author} {\bibinfo {author} {\bibfnamefont {B.}~\bibnamefont
  {Olsthoorn}}, \bibinfo {author} {\bibfnamefont {R.~M.}\ \bibnamefont
  {Geilhufe}}, \bibinfo {author} {\bibfnamefont {S.~S.}\ \bibnamefont
  {Borysov}},\ and\ \bibinfo {author} {\bibfnamefont {A.~V.}\ \bibnamefont
  {Balatsky}},\ }\bibfield  {title} {\bibinfo {title} {Band gap prediction for
  large organic crystal structures with machine learning},\ }\href
  {https://doi.org/https://doi.org/10.1002/qute.201900023} {\bibfield
  {journal} {\bibinfo  {journal} {Advanced Quantum Technologies}\ }\textbf
  {\bibinfo {volume} {2}},\ \bibinfo {pages} {1900023} (\bibinfo {year}
  {2019})}\BibitemShut {NoStop}%
\bibitem [{\citenamefont {Piccinotti}\ \emph {et~al.}(2020)\citenamefont
  {Piccinotti}, \citenamefont {MacDonald}, \citenamefont {Gregory},
  \citenamefont {Youngs},\ and\ \citenamefont {Zheludev}}]{Piccinotti_2020}%
  \BibitemOpen
  \bibfield  {author} {\bibinfo {author} {\bibfnamefont {D.}~\bibnamefont
  {Piccinotti}}, \bibinfo {author} {\bibfnamefont {K.~F.}\ \bibnamefont
  {MacDonald}}, \bibinfo {author} {\bibfnamefont {S.~A.}\ \bibnamefont
  {Gregory}}, \bibinfo {author} {\bibfnamefont {I.}~\bibnamefont {Youngs}},\
  and\ \bibinfo {author} {\bibfnamefont {N.~I.}\ \bibnamefont {Zheludev}},\
  }\bibfield  {title} {\bibinfo {title} {Artificial intelligence for photonics
  and photonic materials},\ }\href {https://doi.org/10.1088/1361-6633/abb4c7}
  {\bibfield  {journal} {\bibinfo  {journal} {Reports on Progress in Physics}\
  }\textbf {\bibinfo {volume} {84}},\ \bibinfo {pages} {012401} (\bibinfo
  {year} {2020})}\BibitemShut {NoStop}%
\bibitem [{\citenamefont {Wiecha}\ \emph {et~al.}(2021)\citenamefont {Wiecha},
  \citenamefont {Arbouet}, \citenamefont {Girard},\ and\ \citenamefont
  {Muskens}}]{Wiecha2020}%
  \BibitemOpen
  \bibfield  {author} {\bibinfo {author} {\bibfnamefont {P.~R.}\ \bibnamefont
  {Wiecha}}, \bibinfo {author} {\bibfnamefont {A.}~\bibnamefont {Arbouet}},
  \bibinfo {author} {\bibfnamefont {C.}~\bibnamefont {Girard}},\ and\ \bibinfo
  {author} {\bibfnamefont {O.~L.}\ \bibnamefont {Muskens}},\ }\bibfield
  {title} {\bibinfo {title} {Deep learning in nano-photonics: inverse design
  and beyond},\ }\href
  {http://www.osapublishing.org/prj/abstract.cfm?URI=prj-9-5-B182} {\bibfield
  {journal} {\bibinfo  {journal} {Photon. Res.}\ }\textbf {\bibinfo {volume}
  {9}},\ \bibinfo {pages} {B182} (\bibinfo {year} {2021})}\BibitemShut
  {NoStop}%
\bibitem [{\citenamefont {Cimini}\ \emph {et~al.}(2020)\citenamefont {Cimini},
  \citenamefont {Barbieri}, \citenamefont {Treps}, \citenamefont {Walschaers},\
  and\ \citenamefont {Parigi}}]{Cimini2020}%
  \BibitemOpen
  \bibfield  {author} {\bibinfo {author} {\bibfnamefont {V.}~\bibnamefont
  {Cimini}}, \bibinfo {author} {\bibfnamefont {M.}~\bibnamefont {Barbieri}},
  \bibinfo {author} {\bibfnamefont {N.}~\bibnamefont {Treps}}, \bibinfo
  {author} {\bibfnamefont {M.}~\bibnamefont {Walschaers}},\ and\ \bibinfo
  {author} {\bibfnamefont {V.}~\bibnamefont {Parigi}},\ }\bibfield  {title}
  {\bibinfo {title} {Neural networks for detecting multimode wigner
  negativity},\ }\href {https://doi.org/10.1103/PhysRevLett.125.160504}
  {\bibfield  {journal} {\bibinfo  {journal} {Phys. Rev. Lett.}\ }\textbf
  {\bibinfo {volume} {125}},\ \bibinfo {pages} {160504} (\bibinfo {year}
  {2020})}\BibitemShut {NoStop}%
\bibitem [{\citenamefont {Kerr}\ \emph {et~al.}(2021)\citenamefont {Kerr},
  \citenamefont {Jose}, \citenamefont {Riggert},\ and\ \citenamefont
  {Mullen}}]{kerr2020automatic}%
  \BibitemOpen
  \bibfield  {author} {\bibinfo {author} {\bibfnamefont {A.}~\bibnamefont
  {Kerr}}, \bibinfo {author} {\bibfnamefont {G.}~\bibnamefont {Jose}}, \bibinfo
  {author} {\bibfnamefont {C.}~\bibnamefont {Riggert}},\ and\ \bibinfo {author}
  {\bibfnamefont {K.}~\bibnamefont {Mullen}},\ }\bibfield  {title} {\bibinfo
  {title} {Automatic learning of topological phase boundaries},\ }\href
  {https://doi.org/10.1103/PhysRevE.103.023310} {\bibfield  {journal} {\bibinfo
   {journal} {Phys. Rev. E}\ }\textbf {\bibinfo {volume} {103}},\ \bibinfo
  {pages} {023310} (\bibinfo {year} {2021})}\BibitemShut {NoStop}%
\bibitem [{\citenamefont {Lidiak}\ and\ \citenamefont
  {Gong}(2020)}]{Lidiak2020}%
  \BibitemOpen
  \bibfield  {author} {\bibinfo {author} {\bibfnamefont {A.}~\bibnamefont
  {Lidiak}}\ and\ \bibinfo {author} {\bibfnamefont {Z.}~\bibnamefont {Gong}},\
  }\bibfield  {title} {\bibinfo {title} {Unsupervised machine learning of
  quantum phase transitions using diffusion maps},\ }\href
  {https://doi.org/10.1103/PhysRevLett.125.225701} {\bibfield  {journal}
  {\bibinfo  {journal} {Phys. Rev. Lett.}\ }\textbf {\bibinfo {volume} {125}},\
  \bibinfo {pages} {225701} (\bibinfo {year} {2020})}\BibitemShut {NoStop}%
\bibitem [{\citenamefont {Cheng}\ \emph {et~al.}(2020)\citenamefont {Cheng},
  \citenamefont {Mazzola}, \citenamefont {Pickard},\ and\ \citenamefont
  {Ceriotti}}]{Cheng2020}%
  \BibitemOpen
  \bibfield  {author} {\bibinfo {author} {\bibfnamefont {B.}~\bibnamefont
  {Cheng}}, \bibinfo {author} {\bibfnamefont {G.}~\bibnamefont {Mazzola}},
  \bibinfo {author} {\bibfnamefont {C.~J.}\ \bibnamefont {Pickard}},\ and\
  \bibinfo {author} {\bibfnamefont {M.}~\bibnamefont {Ceriotti}},\ }\bibfield
  {title} {\bibinfo {title} {Evidence for supercritical behaviour of
  high-pressure liquid hydrogen},\ }\href
  {https://doi.org/10.1038/s41586-020-2677-y} {\bibfield  {journal} {\bibinfo
  {journal} {Nature}\ }\textbf {\bibinfo {volume} {585}},\ \bibinfo {pages}
  {217} (\bibinfo {year} {2020})}\BibitemShut {NoStop}%
\bibitem [{\citenamefont {{Srinivasan}}\ \emph {et~al.}(2020)\citenamefont
  {{Srinivasan}}, \citenamefont {{Batra}}, \citenamefont {{Luo}}, \citenamefont
  {{Loeffler}}, \citenamefont {{Manna}}, \citenamefont {{Chan}}, \citenamefont
  {{Yang}}, \citenamefont {{Yang}}, \citenamefont {{Wen}}, \citenamefont
  {{Darancet}},\ and\ \citenamefont {{Sankaranarayanan}}}]{Argonne2020}%
  \BibitemOpen
  \bibfield  {author} {\bibinfo {author} {\bibfnamefont {S.}~\bibnamefont
  {{Srinivasan}}}, \bibinfo {author} {\bibfnamefont {R.}~\bibnamefont
  {{Batra}}}, \bibinfo {author} {\bibfnamefont {D.}~\bibnamefont {{Luo}}},
  \bibinfo {author} {\bibfnamefont {T.}~\bibnamefont {{Loeffler}}}, \bibinfo
  {author} {\bibfnamefont {S.}~\bibnamefont {{Manna}}}, \bibinfo {author}
  {\bibfnamefont {H.}~\bibnamefont {{Chan}}}, \bibinfo {author} {\bibfnamefont
  {L.}~\bibnamefont {{Yang}}}, \bibinfo {author} {\bibfnamefont
  {W.}~\bibnamefont {{Yang}}}, \bibinfo {author} {\bibfnamefont
  {J.}~\bibnamefont {{Wen}}}, \bibinfo {author} {\bibfnamefont
  {P.}~\bibnamefont {{Darancet}}},\ and\ \bibinfo {author} {\bibfnamefont
  {S.}~\bibnamefont {{Sankaranarayanan}}},\ }\bibfield  {title} {\bibinfo
  {title} {{Machine Learning the Metastable Phase Diagram of Materials}},\
  }\href@noop {} {\bibfield  {journal} {\bibinfo  {journal} {arXiv e-prints}\
  ,\ \bibinfo {eid} {arXiv:2004.08753}} (\bibinfo {year} {2020})},\ \Eprint
  {https://arxiv.org/abs/2004.08753} {arXiv:2004.08753 [cond-mat.mtrl-sci]}
  \BibitemShut {NoStop}%
\bibitem [{\citenamefont {Cross}\ and\ \citenamefont
  {Hohenberg}(1993)}]{Cross_RMP1993}%
  \BibitemOpen
  \bibfield  {author} {\bibinfo {author} {\bibfnamefont {M.~C.}\ \bibnamefont
  {Cross}}\ and\ \bibinfo {author} {\bibfnamefont {P.~C.}\ \bibnamefont
  {Hohenberg}},\ }\bibfield  {title} {\bibinfo {title} {Pattern formation
  outside of equilibrium},\ }\href {https://doi.org/10.1103/RevModPhys.65.851}
  {\bibfield  {journal} {\bibinfo  {journal} {Rev. Mod. Phys.}\ }\textbf
  {\bibinfo {volume} {65}},\ \bibinfo {pages} {851} (\bibinfo {year}
  {1993})}\BibitemShut {NoStop}%
\bibitem [{\citenamefont {Turing}(1952)}]{Turing_RS1952}%
  \BibitemOpen
  \bibfield  {author} {\bibinfo {author} {\bibfnamefont {A.~M.}\ \bibnamefont
  {Turing}},\ }\bibfield  {title} {\bibinfo {title} {The chemical basis of
  morphogenesis},\ }\href {https://doi.org/10.1098/rstb.1952.0012} {\bibfield
  {journal} {\bibinfo  {journal} {Philosophical Transactions of the Royal
  Society of London. Series B, Biological Sciences}\ }\textbf {\bibinfo
  {volume} {237}},\ \bibinfo {pages} {37} (\bibinfo {year} {1952})}\BibitemShut
  {NoStop}%
\bibitem [{\citenamefont {Kibble}(1976)}]{Kibble_JPhysA1976}%
  \BibitemOpen
  \bibfield  {author} {\bibinfo {author} {\bibfnamefont {T.~W.}\ \bibnamefont
  {Kibble}},\ }\bibfield  {title} {\bibinfo {title} {Topology of cosmic domains
  and strings},\ }\href {https://doi.org/10.1088/0305-4470/9/8/029} {\bibfield
  {journal} {\bibinfo  {journal} {Journal of Physics A: Mathematical and
  General}\ }\textbf {\bibinfo {volume} {9}},\ \bibinfo {pages} {1387}
  (\bibinfo {year} {1976})}\BibitemShut {NoStop}%
\bibitem [{\citenamefont {Haken}(1975)}]{Haken1975}%
  \BibitemOpen
  \bibfield  {author} {\bibinfo {author} {\bibfnamefont {H.}~\bibnamefont
  {Haken}},\ }\bibfield  {title} {\bibinfo {title} {Cooperative phenomena in
  systems far from thermal equilibrium and in nonphysical systems},\ }\href
  {https://doi.org/10.1103/RevModPhys.47.67} {\bibfield  {journal} {\bibinfo
  {journal} {Rev. Mod. Phys.}\ }\textbf {\bibinfo {volume} {47}},\ \bibinfo
  {pages} {67} (\bibinfo {year} {1975})}\BibitemShut {NoStop}%
\bibitem [{\citenamefont {Haken}\ and\ \citenamefont {Ohno}(1978)}]{Haken1978}%
  \BibitemOpen
  \bibfield  {author} {\bibinfo {author} {\bibfnamefont {H.}~\bibnamefont
  {Haken}}\ and\ \bibinfo {author} {\bibfnamefont {H.}~\bibnamefont {Ohno}},\
  }\bibfield  {title} {\bibinfo {title} {Onset of ultrashort laser pulses:
  First or second order phase transition?},\ }\href
  {https://doi.org/https://doi.org/10.1016/0030-4018(78)90357-7} {\bibfield
  {journal} {\bibinfo  {journal} {Optics Communications}\ }\textbf {\bibinfo
  {volume} {26}},\ \bibinfo {pages} {117} (\bibinfo {year} {1978})}\BibitemShut
  {NoStop}%
\bibitem [{\citenamefont {Haken}(1978)}]{HakenBook}%
  \BibitemOpen
  \bibfield  {author} {\bibinfo {author} {\bibfnamefont {H.}~\bibnamefont
  {Haken}},\ }\href {https://doi.org/10.1007/978-3-642-96469-5} {\emph
  {\bibinfo {title} {Synergetics, an Introduction: Nonequilibrium Phase
  Transitions and Self-Organization in Physics, Chemistry, and Biology}}},\
  Vol.~\bibinfo {volume} {1}\ (\bibinfo  {publisher} {Springer-Verlag Berlin
  Heidelberg},\ \bibinfo {year} {1978})\BibitemShut {NoStop}%
\bibitem [{\citenamefont {Fruchart}\ \emph {et~al.}(2021)\citenamefont
  {Fruchart}, \citenamefont {Hanai}, \citenamefont {Littlewood},\ and\
  \citenamefont {Vitelli}}]{Fruchart_Nature2021}%
  \BibitemOpen
  \bibfield  {author} {\bibinfo {author} {\bibfnamefont {M.}~\bibnamefont
  {Fruchart}}, \bibinfo {author} {\bibfnamefont {R.}~\bibnamefont {Hanai}},
  \bibinfo {author} {\bibfnamefont {P.~B.}\ \bibnamefont {Littlewood}},\ and\
  \bibinfo {author} {\bibfnamefont {V.}~\bibnamefont {Vitelli}},\ }\bibfield
  {title} {\bibinfo {title} {Non-reciprocal phase transitions},\ }\href
  {https://doi.org/10.1038/s41586-021-03375-9} {\bibfield  {journal} {\bibinfo
  {journal} {Nature}\ }\textbf {\bibinfo {volume} {592}},\ \bibinfo {pages}
  {363} (\bibinfo {year} {2021})}\BibitemShut {NoStop}%
\bibitem [{\citenamefont {Baumann}\ \emph {et~al.}(2010)\citenamefont
  {Baumann}, \citenamefont {Guerlin}, \citenamefont {Brennecke},\ and\
  \citenamefont {Esslinger}}]{Baumann_Nature2010}%
  \BibitemOpen
  \bibfield  {author} {\bibinfo {author} {\bibfnamefont {K.}~\bibnamefont
  {Baumann}}, \bibinfo {author} {\bibfnamefont {C.}~\bibnamefont {Guerlin}},
  \bibinfo {author} {\bibfnamefont {F.}~\bibnamefont {Brennecke}},\ and\
  \bibinfo {author} {\bibfnamefont {T.}~\bibnamefont {Esslinger}},\ }\bibfield
  {title} {\bibinfo {title} {Dicke quantum phase transition with a superfluid
  gas in an optical cavity},\ }\href {https://doi.org/10.1038/nature09009}
  {\bibfield  {journal} {\bibinfo  {journal} {Nature}\ }\textbf {\bibinfo
  {volume} {464}},\ \bibinfo {pages} {1301} (\bibinfo {year}
  {2010})}\BibitemShut {NoStop}%
\bibitem [{\citenamefont {Kavokin}\ \emph {et~al.}(2017)\citenamefont
  {Kavokin}, \citenamefont {Baumberg}, \citenamefont {Malpuech},\ and\
  \citenamefont {Laussy}}]{KavokinBook}%
  \BibitemOpen
  \bibfield  {author} {\bibinfo {author} {\bibfnamefont {A.~V.}\ \bibnamefont
  {Kavokin}}, \bibinfo {author} {\bibfnamefont {J.~J.}\ \bibnamefont
  {Baumberg}}, \bibinfo {author} {\bibfnamefont {G.}~\bibnamefont {Malpuech}},\
  and\ \bibinfo {author} {\bibfnamefont {F.~P.}\ \bibnamefont {Laussy}},\
  }\href@noop {} {\emph {\bibinfo {title} {Microcavities}}},\ Vol.~\bibinfo
  {volume} {21}\ (\bibinfo  {publisher} {Oxford university press},\ \bibinfo
  {year} {2017})\BibitemShut {NoStop}%
\bibitem [{\citenamefont {Carusotto}\ and\ \citenamefont
  {Ciuti}(2013)}]{CarusottoCiutiRev}%
  \BibitemOpen
  \bibfield  {author} {\bibinfo {author} {\bibfnamefont {I.}~\bibnamefont
  {Carusotto}}\ and\ \bibinfo {author} {\bibfnamefont {C.}~\bibnamefont
  {Ciuti}},\ }\bibfield  {title} {\bibinfo {title} {Quantum fluids of light},\
  }\href {https://doi.org/10.1103/RevModPhys.85.299} {\bibfield  {journal}
  {\bibinfo  {journal} {Rev. Mod. Phys.}\ }\textbf {\bibinfo {volume} {85}},\
  \bibinfo {pages} {299} (\bibinfo {year} {2013})}\BibitemShut {NoStop}%
\bibitem [{\citenamefont {Kasprzak}\ \emph {et~al.}(2006)\citenamefont
  {Kasprzak}, \citenamefont {Richard}, \citenamefont {Kundermann},
  \citenamefont {Baas}, \citenamefont {Jeambrun}, \citenamefont {Keeling},
  \citenamefont {Marchetti}, \citenamefont {Szyma{\'{n}}ska}, \citenamefont
  {Andr{\'e}}, \citenamefont {Staehli}, \citenamefont {Savona}, \citenamefont
  {Littlewood}, \citenamefont {Deveaud},\ and\ \citenamefont
  {Dang}}]{Kasprzak_Nature2006}%
  \BibitemOpen
  \bibfield  {author} {\bibinfo {author} {\bibfnamefont {J.}~\bibnamefont
  {Kasprzak}}, \bibinfo {author} {\bibfnamefont {M.}~\bibnamefont {Richard}},
  \bibinfo {author} {\bibfnamefont {S.}~\bibnamefont {Kundermann}}, \bibinfo
  {author} {\bibfnamefont {A.}~\bibnamefont {Baas}}, \bibinfo {author}
  {\bibfnamefont {P.}~\bibnamefont {Jeambrun}}, \bibinfo {author}
  {\bibfnamefont {J.~M.~J.}\ \bibnamefont {Keeling}}, \bibinfo {author}
  {\bibfnamefont {F.~M.}\ \bibnamefont {Marchetti}}, \bibinfo {author}
  {\bibfnamefont {M.~H.}\ \bibnamefont {Szyma{\'{n}}ska}}, \bibinfo {author}
  {\bibfnamefont {R.}~\bibnamefont {Andr{\'e}}}, \bibinfo {author}
  {\bibfnamefont {J.~L.}\ \bibnamefont {Staehli}}, \bibinfo {author}
  {\bibfnamefont {V.}~\bibnamefont {Savona}}, \bibinfo {author} {\bibfnamefont
  {P.~B.}\ \bibnamefont {Littlewood}}, \bibinfo {author} {\bibfnamefont
  {B.}~\bibnamefont {Deveaud}},\ and\ \bibinfo {author} {\bibfnamefont {L.~S.}\
  \bibnamefont {Dang}},\ }\bibfield  {title} {\bibinfo {title} {Bose--einstein
  condensation of exciton polaritons},\ }\href
  {https://doi.org/10.1038/nature05131} {\bibfield  {journal} {\bibinfo
  {journal} {Nature}\ }\textbf {\bibinfo {volume} {443}},\ \bibinfo {pages}
  {409} (\bibinfo {year} {2006})}\BibitemShut {NoStop}%
\bibitem [{\citenamefont {Christopoulos}\ \emph {et~al.}(2007)\citenamefont
  {Christopoulos}, \citenamefont {von H\"ogersthal}, \citenamefont {Grundy},
  \citenamefont {Lagoudakis}, \citenamefont {Kavokin}, \citenamefont
  {Baumberg}, \citenamefont {Christmann}, \citenamefont {Butt\'e},
  \citenamefont {Feltin}, \citenamefont {Carlin},\ and\ \citenamefont
  {Grandjean}}]{Christopoulos_PRL2007}%
  \BibitemOpen
  \bibfield  {author} {\bibinfo {author} {\bibfnamefont {S.}~\bibnamefont
  {Christopoulos}}, \bibinfo {author} {\bibfnamefont {G.~B.~H.}\ \bibnamefont
  {von H\"ogersthal}}, \bibinfo {author} {\bibfnamefont {A.~J.~D.}\
  \bibnamefont {Grundy}}, \bibinfo {author} {\bibfnamefont {P.~G.}\
  \bibnamefont {Lagoudakis}}, \bibinfo {author} {\bibfnamefont {A.~V.}\
  \bibnamefont {Kavokin}}, \bibinfo {author} {\bibfnamefont {J.~J.}\
  \bibnamefont {Baumberg}}, \bibinfo {author} {\bibfnamefont {G.}~\bibnamefont
  {Christmann}}, \bibinfo {author} {\bibfnamefont {R.}~\bibnamefont {Butt\'e}},
  \bibinfo {author} {\bibfnamefont {E.}~\bibnamefont {Feltin}}, \bibinfo
  {author} {\bibfnamefont {J.-F.}\ \bibnamefont {Carlin}},\ and\ \bibinfo
  {author} {\bibfnamefont {N.}~\bibnamefont {Grandjean}},\ }\bibfield  {title}
  {\bibinfo {title} {Room-temperature polariton lasing in semiconductor
  microcavities},\ }\href {https://doi.org/10.1103/PhysRevLett.98.126405}
  {\bibfield  {journal} {\bibinfo  {journal} {Phys. Rev. Lett.}\ }\textbf
  {\bibinfo {volume} {98}},\ \bibinfo {pages} {126405} (\bibinfo {year}
  {2007})}\BibitemShut {NoStop}%
\bibitem [{\citenamefont {Schneider}\ \emph {et~al.}(2013)\citenamefont
  {Schneider}, \citenamefont {Rahimi-Iman}, \citenamefont {Kim}, \citenamefont
  {Fischer}, \citenamefont {Savenko}, \citenamefont {Amthor}, \citenamefont
  {Lermer}, \citenamefont {Wolf}, \citenamefont {Worschech}, \citenamefont
  {Kulakovskii}, \citenamefont {Shelykh}, \citenamefont {Kamp}, \citenamefont
  {Reitzenstein}, \citenamefont {Forchel}, \citenamefont {Yamamoto},\ and\
  \citenamefont {H{\"o}fling}}]{Schneider2013}%
  \BibitemOpen
  \bibfield  {author} {\bibinfo {author} {\bibfnamefont {C.}~\bibnamefont
  {Schneider}}, \bibinfo {author} {\bibfnamefont {A.}~\bibnamefont
  {Rahimi-Iman}}, \bibinfo {author} {\bibfnamefont {N.~Y.}\ \bibnamefont
  {Kim}}, \bibinfo {author} {\bibfnamefont {J.}~\bibnamefont {Fischer}},
  \bibinfo {author} {\bibfnamefont {I.~G.}\ \bibnamefont {Savenko}}, \bibinfo
  {author} {\bibfnamefont {M.}~\bibnamefont {Amthor}}, \bibinfo {author}
  {\bibfnamefont {M.}~\bibnamefont {Lermer}}, \bibinfo {author} {\bibfnamefont
  {A.}~\bibnamefont {Wolf}}, \bibinfo {author} {\bibfnamefont {L.}~\bibnamefont
  {Worschech}}, \bibinfo {author} {\bibfnamefont {V.~D.}\ \bibnamefont
  {Kulakovskii}}, \bibinfo {author} {\bibfnamefont {I.~A.}\ \bibnamefont
  {Shelykh}}, \bibinfo {author} {\bibfnamefont {M.}~\bibnamefont {Kamp}},
  \bibinfo {author} {\bibfnamefont {S.}~\bibnamefont {Reitzenstein}}, \bibinfo
  {author} {\bibfnamefont {A.}~\bibnamefont {Forchel}}, \bibinfo {author}
  {\bibfnamefont {Y.}~\bibnamefont {Yamamoto}},\ and\ \bibinfo {author}
  {\bibfnamefont {S.}~\bibnamefont {H{\"o}fling}},\ }\bibfield  {title}
  {\bibinfo {title} {An electrically pumped polariton laser},\ }\href
  {https://doi.org/10.1038/nature12036} {\bibfield  {journal} {\bibinfo
  {journal} {Nature}\ }\textbf {\bibinfo {volume} {497}},\ \bibinfo {pages}
  {348} (\bibinfo {year} {2013})}\BibitemShut {NoStop}%
\bibitem [{\citenamefont {Leyder}\ \emph {et~al.}(2007)\citenamefont {Leyder},
  \citenamefont {Romanelli}, \citenamefont {Karr}, \citenamefont {Giacobino},
  \citenamefont {Liew}, \citenamefont {Glazov}, \citenamefont {Kavokin},
  \citenamefont {Malpuech},\ and\ \citenamefont
  {Bramati}}]{Leyder_NatPhys2007}%
  \BibitemOpen
  \bibfield  {author} {\bibinfo {author} {\bibfnamefont {C.}~\bibnamefont
  {Leyder}}, \bibinfo {author} {\bibfnamefont {M.}~\bibnamefont {Romanelli}},
  \bibinfo {author} {\bibfnamefont {J.~P.}\ \bibnamefont {Karr}}, \bibinfo
  {author} {\bibfnamefont {E.}~\bibnamefont {Giacobino}}, \bibinfo {author}
  {\bibfnamefont {T.~C.~H.}\ \bibnamefont {Liew}}, \bibinfo {author}
  {\bibfnamefont {M.~M.}\ \bibnamefont {Glazov}}, \bibinfo {author}
  {\bibfnamefont {A.~V.}\ \bibnamefont {Kavokin}}, \bibinfo {author}
  {\bibfnamefont {G.}~\bibnamefont {Malpuech}},\ and\ \bibinfo {author}
  {\bibfnamefont {A.}~\bibnamefont {Bramati}},\ }\bibfield  {title} {\bibinfo
  {title} {Observation of the optical spin hall effect},\ }\href
  {https://doi.org/10.1038/nphys676} {\bibfield  {journal} {\bibinfo  {journal}
  {Nature Physics}\ }\textbf {\bibinfo {volume} {3}},\ \bibinfo {pages} {628}
  (\bibinfo {year} {2007})}\BibitemShut {NoStop}%
\bibitem [{\citenamefont {Chana}\ \emph {et~al.}(2015)\citenamefont {Chana},
  \citenamefont {Sich}, \citenamefont {Fras}, \citenamefont {Gorbach},
  \citenamefont {Skryabin}, \citenamefont {Cancellieri}, \citenamefont
  {Cerda-M\'endez}, \citenamefont {Biermann}, \citenamefont {Hey},
  \citenamefont {Santos}, \citenamefont {Skolnick},\ and\ \citenamefont
  {Krizhanovskii}}]{Chana2015}%
  \BibitemOpen
  \bibfield  {author} {\bibinfo {author} {\bibfnamefont {J.~K.}\ \bibnamefont
  {Chana}}, \bibinfo {author} {\bibfnamefont {M.}~\bibnamefont {Sich}},
  \bibinfo {author} {\bibfnamefont {F.}~\bibnamefont {Fras}}, \bibinfo {author}
  {\bibfnamefont {A.~V.}\ \bibnamefont {Gorbach}}, \bibinfo {author}
  {\bibfnamefont {D.~V.}\ \bibnamefont {Skryabin}}, \bibinfo {author}
  {\bibfnamefont {E.}~\bibnamefont {Cancellieri}}, \bibinfo {author}
  {\bibfnamefont {E.~A.}\ \bibnamefont {Cerda-M\'endez}}, \bibinfo {author}
  {\bibfnamefont {K.}~\bibnamefont {Biermann}}, \bibinfo {author}
  {\bibfnamefont {R.}~\bibnamefont {Hey}}, \bibinfo {author} {\bibfnamefont
  {P.~V.}\ \bibnamefont {Santos}}, \bibinfo {author} {\bibfnamefont {M.~S.}\
  \bibnamefont {Skolnick}},\ and\ \bibinfo {author} {\bibfnamefont {D.~N.}\
  \bibnamefont {Krizhanovskii}},\ }\bibfield  {title} {\bibinfo {title}
  {Spatial patterns of dissipative polariton solitons in semiconductor
  microcavities},\ }\href {https://doi.org/10.1103/PhysRevLett.115.256401}
  {\bibfield  {journal} {\bibinfo  {journal} {Phys. Rev. Lett.}\ }\textbf
  {\bibinfo {volume} {115}},\ \bibinfo {pages} {256401} (\bibinfo {year}
  {2015})}\BibitemShut {NoStop}%
\bibitem [{\citenamefont {Lagoudakis}\ \emph {et~al.}(2008)\citenamefont
  {Lagoudakis}, \citenamefont {Wouters}, \citenamefont {Richard}, \citenamefont
  {Baas}, \citenamefont {Carusotto}, \citenamefont {Andr{\'e}}, \citenamefont
  {Dang},\ and\ \citenamefont {Deveaud-Pl{\'e}dran}}]{Lagoudakis2008}%
  \BibitemOpen
  \bibfield  {author} {\bibinfo {author} {\bibfnamefont {K.~G.}\ \bibnamefont
  {Lagoudakis}}, \bibinfo {author} {\bibfnamefont {M.}~\bibnamefont {Wouters}},
  \bibinfo {author} {\bibfnamefont {M.}~\bibnamefont {Richard}}, \bibinfo
  {author} {\bibfnamefont {A.}~\bibnamefont {Baas}}, \bibinfo {author}
  {\bibfnamefont {I.}~\bibnamefont {Carusotto}}, \bibinfo {author}
  {\bibfnamefont {R.}~\bibnamefont {Andr{\'e}}}, \bibinfo {author}
  {\bibfnamefont {L.~S.}\ \bibnamefont {Dang}},\ and\ \bibinfo {author}
  {\bibfnamefont {B.}~\bibnamefont {Deveaud-Pl{\'e}dran}},\ }\bibfield  {title}
  {\bibinfo {title} {Quantized vortices in an exciton--polariton condensate},\
  }\href {https://doi.org/10.1038/nphys1051} {\bibfield  {journal} {\bibinfo
  {journal} {Nature Physics}\ }\textbf {\bibinfo {volume} {4}},\ \bibinfo
  {pages} {706} (\bibinfo {year} {2008})}\BibitemShut {NoStop}%
\bibitem [{\citenamefont {Caputo}\ \emph {et~al.}(2019)\citenamefont {Caputo},
  \citenamefont {Bobrovska}, \citenamefont {Ballarini}, \citenamefont
  {Matuszewski}, \citenamefont {De~Giorgi}, \citenamefont {Dominici},
  \citenamefont {West}, \citenamefont {Pfeiffer}, \citenamefont {Gigli},\ and\
  \citenamefont {Sanvitto}}]{Caputo2019}%
  \BibitemOpen
  \bibfield  {author} {\bibinfo {author} {\bibfnamefont {D.}~\bibnamefont
  {Caputo}}, \bibinfo {author} {\bibfnamefont {N.}~\bibnamefont {Bobrovska}},
  \bibinfo {author} {\bibfnamefont {D.}~\bibnamefont {Ballarini}}, \bibinfo
  {author} {\bibfnamefont {M.}~\bibnamefont {Matuszewski}}, \bibinfo {author}
  {\bibfnamefont {M.}~\bibnamefont {De~Giorgi}}, \bibinfo {author}
  {\bibfnamefont {L.}~\bibnamefont {Dominici}}, \bibinfo {author}
  {\bibfnamefont {K.}~\bibnamefont {West}}, \bibinfo {author} {\bibfnamefont
  {L.~N.}\ \bibnamefont {Pfeiffer}}, \bibinfo {author} {\bibfnamefont
  {G.}~\bibnamefont {Gigli}},\ and\ \bibinfo {author} {\bibfnamefont
  {D.}~\bibnamefont {Sanvitto}},\ }\bibfield  {title} {\bibinfo {title}
  {Josephson vortices induced by phase twisting a polariton superfluid},\
  }\href {https://doi.org/10.1038/s41566-019-0425-3} {\bibfield  {journal}
  {\bibinfo  {journal} {Nature Photonics}\ }\textbf {\bibinfo {volume} {13}},\
  \bibinfo {pages} {488} (\bibinfo {year} {2019})}\BibitemShut {NoStop}%
\bibitem [{\citenamefont {Delteil}\ \emph {et~al.}(2019)\citenamefont
  {Delteil}, \citenamefont {Fink}, \citenamefont {Schade}, \citenamefont
  {H{\"o}fling}, \citenamefont {Schneider},\ and\ \citenamefont
  {{\.{I}}mamo{\u{g}}lu}}]{Delteil2019}%
  \BibitemOpen
  \bibfield  {author} {\bibinfo {author} {\bibfnamefont {A.}~\bibnamefont
  {Delteil}}, \bibinfo {author} {\bibfnamefont {T.}~\bibnamefont {Fink}},
  \bibinfo {author} {\bibfnamefont {A.}~\bibnamefont {Schade}}, \bibinfo
  {author} {\bibfnamefont {S.}~\bibnamefont {H{\"o}fling}}, \bibinfo {author}
  {\bibfnamefont {C.}~\bibnamefont {Schneider}},\ and\ \bibinfo {author}
  {\bibfnamefont {A.}~\bibnamefont {{\.{I}}mamo{\u{g}}lu}},\ }\bibfield
  {title} {\bibinfo {title} {Towards polariton blockade of confined
  exciton--polaritons},\ }\href {https://doi.org/10.1038/s41563-019-0282-y}
  {\bibfield  {journal} {\bibinfo  {journal} {Nature Materials}\ }\textbf
  {\bibinfo {volume} {18}},\ \bibinfo {pages} {219} (\bibinfo {year}
  {2019})}\BibitemShut {NoStop}%
\bibitem [{\citenamefont {Schneider}\ \emph {et~al.}(2016)\citenamefont
  {Schneider}, \citenamefont {Winkler}, \citenamefont {Fraser}, \citenamefont
  {Kamp}, \citenamefont {Yamamoto}, \citenamefont {Ostrovskaya},\ and\
  \citenamefont {Höfling}}]{Schneider_RepProgPhys2017}%
  \BibitemOpen
  \bibfield  {author} {\bibinfo {author} {\bibfnamefont {C.}~\bibnamefont
  {Schneider}}, \bibinfo {author} {\bibfnamefont {K.}~\bibnamefont {Winkler}},
  \bibinfo {author} {\bibfnamefont {M.~D.}\ \bibnamefont {Fraser}}, \bibinfo
  {author} {\bibfnamefont {M.}~\bibnamefont {Kamp}}, \bibinfo {author}
  {\bibfnamefont {Y.}~\bibnamefont {Yamamoto}}, \bibinfo {author}
  {\bibfnamefont {E.~A.}\ \bibnamefont {Ostrovskaya}},\ and\ \bibinfo {author}
  {\bibfnamefont {S.}~\bibnamefont {Höfling}},\ }\bibfield  {title} {\bibinfo
  {title} {Exciton-polariton trapping and potential landscape engineering},\
  }\href {https://doi.org/10.1088/0034-4885/80/1/016503} {\bibfield  {journal}
  {\bibinfo  {journal} {Reports on Progress in Physics}\ }\textbf {\bibinfo
  {volume} {80}},\ \bibinfo {pages} {016503} (\bibinfo {year}
  {2016})}\BibitemShut {NoStop}%
\bibitem [{\citenamefont {Jacqmin}\ \emph {et~al.}(2014)\citenamefont
  {Jacqmin}, \citenamefont {Carusotto}, \citenamefont {Sagnes}, \citenamefont
  {Abbarchi}, \citenamefont {Solnyshkov}, \citenamefont {Malpuech},
  \citenamefont {Galopin}, \citenamefont {Lema\^{\i}tre}, \citenamefont
  {Bloch},\ and\ \citenamefont {Amo}}]{Jacqmin2014}%
  \BibitemOpen
  \bibfield  {author} {\bibinfo {author} {\bibfnamefont {T.}~\bibnamefont
  {Jacqmin}}, \bibinfo {author} {\bibfnamefont {I.}~\bibnamefont {Carusotto}},
  \bibinfo {author} {\bibfnamefont {I.}~\bibnamefont {Sagnes}}, \bibinfo
  {author} {\bibfnamefont {M.}~\bibnamefont {Abbarchi}}, \bibinfo {author}
  {\bibfnamefont {D.~D.}\ \bibnamefont {Solnyshkov}}, \bibinfo {author}
  {\bibfnamefont {G.}~\bibnamefont {Malpuech}}, \bibinfo {author}
  {\bibfnamefont {E.}~\bibnamefont {Galopin}}, \bibinfo {author} {\bibfnamefont
  {A.}~\bibnamefont {Lema\^{\i}tre}}, \bibinfo {author} {\bibfnamefont
  {J.}~\bibnamefont {Bloch}},\ and\ \bibinfo {author} {\bibfnamefont
  {A.}~\bibnamefont {Amo}},\ }\bibfield  {title} {\bibinfo {title} {Direct
  observation of dirac cones and a flatband in a honeycomb lattice for
  polaritons},\ }\href {https://doi.org/10.1103/PhysRevLett.112.116402}
  {\bibfield  {journal} {\bibinfo  {journal} {Phys. Rev. Lett.}\ }\textbf
  {\bibinfo {volume} {112}},\ \bibinfo {pages} {116402} (\bibinfo {year}
  {2014})}\BibitemShut {NoStop}%
\bibitem [{\citenamefont {Jayaprakash}\ \emph {et~al.}(2020)\citenamefont
  {Jayaprakash}, \citenamefont {Whittaker}, \citenamefont {Georgiou},
  \citenamefont {Game}, \citenamefont {McGhee}, \citenamefont {Coles},\ and\
  \citenamefont {Lidzey}}]{Jayaprakash_ACS2020}%
  \BibitemOpen
  \bibfield  {author} {\bibinfo {author} {\bibfnamefont {R.}~\bibnamefont
  {Jayaprakash}}, \bibinfo {author} {\bibfnamefont {C.~E.}\ \bibnamefont
  {Whittaker}}, \bibinfo {author} {\bibfnamefont {K.}~\bibnamefont {Georgiou}},
  \bibinfo {author} {\bibfnamefont {O.~S.}\ \bibnamefont {Game}}, \bibinfo
  {author} {\bibfnamefont {K.~E.}\ \bibnamefont {McGhee}}, \bibinfo {author}
  {\bibfnamefont {D.~M.}\ \bibnamefont {Coles}},\ and\ \bibinfo {author}
  {\bibfnamefont {D.~G.}\ \bibnamefont {Lidzey}},\ }\bibfield  {title}
  {\bibinfo {title} {Two-dimensional organic-exciton polariton lattice
  fabricated using laser patterning},\ }\href
  {https://doi.org/10.1021/acsphotonics.0c00867} {\bibfield  {journal}
  {\bibinfo  {journal} {ACS Photonics}\ }\textbf {\bibinfo {volume} {7}},\
  \bibinfo {pages} {2273} (\bibinfo {year} {2020})}\BibitemShut {NoStop}%
\bibitem [{\citenamefont {Wertz}\ \emph {et~al.}(2010)\citenamefont {Wertz},
  \citenamefont {Ferrier}, \citenamefont {Solnyshkov}, \citenamefont {Johne},
  \citenamefont {Sanvitto}, \citenamefont {Lema{\^i}tre}, \citenamefont
  {Sagnes}, \citenamefont {Grousson}, \citenamefont {Kavokin}, \citenamefont
  {Senellart}, \citenamefont {Malpuech},\ and\ \citenamefont
  {Bloch}}]{Wertz_NatPhys2010}%
  \BibitemOpen
  \bibfield  {author} {\bibinfo {author} {\bibfnamefont {E.}~\bibnamefont
  {Wertz}}, \bibinfo {author} {\bibfnamefont {L.}~\bibnamefont {Ferrier}},
  \bibinfo {author} {\bibfnamefont {D.~D.}\ \bibnamefont {Solnyshkov}},
  \bibinfo {author} {\bibfnamefont {R.}~\bibnamefont {Johne}}, \bibinfo
  {author} {\bibfnamefont {D.}~\bibnamefont {Sanvitto}}, \bibinfo {author}
  {\bibfnamefont {A.}~\bibnamefont {Lema{\^i}tre}}, \bibinfo {author}
  {\bibfnamefont {I.}~\bibnamefont {Sagnes}}, \bibinfo {author} {\bibfnamefont
  {R.}~\bibnamefont {Grousson}}, \bibinfo {author} {\bibfnamefont {A.~V.}\
  \bibnamefont {Kavokin}}, \bibinfo {author} {\bibfnamefont {P.}~\bibnamefont
  {Senellart}}, \bibinfo {author} {\bibfnamefont {G.}~\bibnamefont
  {Malpuech}},\ and\ \bibinfo {author} {\bibfnamefont {J.}~\bibnamefont
  {Bloch}},\ }\bibfield  {title} {\bibinfo {title} {Spontaneous formation and
  optical manipulation of extended polariton condensates},\ }\href
  {https://doi.org/10.1038/nphys1750} {\bibfield  {journal} {\bibinfo
  {journal} {Nature Physics}\ }\textbf {\bibinfo {volume} {6}},\ \bibinfo
  {pages} {860} (\bibinfo {year} {2010})}\BibitemShut {NoStop}%
\bibitem [{\citenamefont {Pickup}\ \emph {et~al.}(2020)\citenamefont {Pickup},
  \citenamefont {Sigurdsson}, \citenamefont {Ruostekoski},\ and\ \citenamefont
  {Lagoudakis}}]{Pickup_NatComm2020}%
  \BibitemOpen
  \bibfield  {author} {\bibinfo {author} {\bibfnamefont {L.}~\bibnamefont
  {Pickup}}, \bibinfo {author} {\bibfnamefont {H.}~\bibnamefont {Sigurdsson}},
  \bibinfo {author} {\bibfnamefont {J.}~\bibnamefont {Ruostekoski}},\ and\
  \bibinfo {author} {\bibfnamefont {P.~G.}\ \bibnamefont {Lagoudakis}},\
  }\bibfield  {title} {\bibinfo {title} {Synthetic band-structure engineering
  in polariton crystals with non-hermitian topological phases},\ }\href
  {https://doi.org/10.1038/s41467-020-18213-1} {\bibfield  {journal} {\bibinfo
  {journal} {Nature Communications}\ }\textbf {\bibinfo {volume} {11}},\
  \bibinfo {pages} {4431} (\bibinfo {year} {2020})}\BibitemShut {NoStop}%
\bibitem [{\citenamefont {T\"{o}pfer}\ \emph {et~al.}(2021)\citenamefont
  {T\"{o}pfer}, \citenamefont {Chatzopoulos}, \citenamefont {Sigurdsson},
  \citenamefont {Cookson}, \citenamefont {Rubo},\ and\ \citenamefont
  {Lagoudakis}}]{Topfer_Optica2020}%
  \BibitemOpen
  \bibfield  {author} {\bibinfo {author} {\bibfnamefont {J.~D.}\ \bibnamefont
  {T\"{o}pfer}}, \bibinfo {author} {\bibfnamefont {I.}~\bibnamefont
  {Chatzopoulos}}, \bibinfo {author} {\bibfnamefont {H.}~\bibnamefont
  {Sigurdsson}}, \bibinfo {author} {\bibfnamefont {T.}~\bibnamefont {Cookson}},
  \bibinfo {author} {\bibfnamefont {Y.~G.}\ \bibnamefont {Rubo}},\ and\
  \bibinfo {author} {\bibfnamefont {P.~G.}\ \bibnamefont {Lagoudakis}},\
  }\bibfield  {title} {\bibinfo {title} {Engineering spatial coherence in
  lattices of polariton condensates},\ }\href
  {https://doi.org/10.1364/OPTICA.409976} {\bibfield  {journal} {\bibinfo
  {journal} {Optica}\ }\textbf {\bibinfo {volume} {8}},\ \bibinfo {pages} {106}
  (\bibinfo {year} {2021})}\BibitemShut {NoStop}%
\bibitem [{\citenamefont {Askitopoulos}\ \emph {et~al.}(2013)\citenamefont
  {Askitopoulos}, \citenamefont {Ohadi}, \citenamefont {Kavokin}, \citenamefont
  {Hatzopoulos}, \citenamefont {Savvidis},\ and\ \citenamefont
  {Lagoudakis}}]{Askitopoulos2013}%
  \BibitemOpen
  \bibfield  {author} {\bibinfo {author} {\bibfnamefont {A.}~\bibnamefont
  {Askitopoulos}}, \bibinfo {author} {\bibfnamefont {H.}~\bibnamefont {Ohadi}},
  \bibinfo {author} {\bibfnamefont {A.~V.}\ \bibnamefont {Kavokin}}, \bibinfo
  {author} {\bibfnamefont {Z.}~\bibnamefont {Hatzopoulos}}, \bibinfo {author}
  {\bibfnamefont {P.~G.}\ \bibnamefont {Savvidis}},\ and\ \bibinfo {author}
  {\bibfnamefont {P.~G.}\ \bibnamefont {Lagoudakis}},\ }\bibfield  {title}
  {\bibinfo {title} {Polariton condensation in an optically induced
  two-dimensional potential},\ }\href
  {https://doi.org/10.1103/PhysRevB.88.041308} {\bibfield  {journal} {\bibinfo
  {journal} {Phys. Rev. B}\ }\textbf {\bibinfo {volume} {88}},\ \bibinfo
  {pages} {041308} (\bibinfo {year} {2013})}\BibitemShut {NoStop}%
\bibitem [{\citenamefont {Ohadi}\ \emph {et~al.}(2017)\citenamefont {Ohadi},
  \citenamefont {Ramsay}, \citenamefont {Sigurdsson}, \citenamefont {del
  Valle-Inclan~Redondo}, \citenamefont {Tsintzos}, \citenamefont {Hatzopoulos},
  \citenamefont {Liew}, \citenamefont {Shelykh}, \citenamefont {Rubo},
  \citenamefont {Savvidis},\ and\ \citenamefont {Baumberg}}]{Ohadi2017}%
  \BibitemOpen
  \bibfield  {author} {\bibinfo {author} {\bibfnamefont {H.}~\bibnamefont
  {Ohadi}}, \bibinfo {author} {\bibfnamefont {A.~J.}\ \bibnamefont {Ramsay}},
  \bibinfo {author} {\bibfnamefont {H.}~\bibnamefont {Sigurdsson}}, \bibinfo
  {author} {\bibfnamefont {Y.}~\bibnamefont {del Valle-Inclan~Redondo}},
  \bibinfo {author} {\bibfnamefont {S.~I.}\ \bibnamefont {Tsintzos}}, \bibinfo
  {author} {\bibfnamefont {Z.}~\bibnamefont {Hatzopoulos}}, \bibinfo {author}
  {\bibfnamefont {T.~C.~H.}\ \bibnamefont {Liew}}, \bibinfo {author}
  {\bibfnamefont {I.~A.}\ \bibnamefont {Shelykh}}, \bibinfo {author}
  {\bibfnamefont {Y.~G.}\ \bibnamefont {Rubo}}, \bibinfo {author}
  {\bibfnamefont {P.~G.}\ \bibnamefont {Savvidis}},\ and\ \bibinfo {author}
  {\bibfnamefont {J.~J.}\ \bibnamefont {Baumberg}},\ }\bibfield  {title}
  {\bibinfo {title} {Spin order and phase transitions in chains of polariton
  condensates},\ }\href {https://doi.org/10.1103/PhysRevLett.119.067401}
  {\bibfield  {journal} {\bibinfo  {journal} {Phys. Rev. Lett.}\ }\textbf
  {\bibinfo {volume} {119}},\ \bibinfo {pages} {067401} (\bibinfo {year}
  {2017})}\BibitemShut {NoStop}%
\bibitem [{\citenamefont {Ohadi}\ \emph {et~al.}(2018)\citenamefont {Ohadi},
  \citenamefont {del Valle-Inclan~Redondo}, \citenamefont {Ramsay},
  \citenamefont {Hatzopoulos}, \citenamefont {Liew}, \citenamefont {Eastham},
  \citenamefont {Savvidis},\ and\ \citenamefont {Baumberg}}]{Ohadi_PRB2018}%
  \BibitemOpen
  \bibfield  {author} {\bibinfo {author} {\bibfnamefont {H.}~\bibnamefont
  {Ohadi}}, \bibinfo {author} {\bibfnamefont {Y.}~\bibnamefont {del
  Valle-Inclan~Redondo}}, \bibinfo {author} {\bibfnamefont {A.~J.}\
  \bibnamefont {Ramsay}}, \bibinfo {author} {\bibfnamefont {Z.}~\bibnamefont
  {Hatzopoulos}}, \bibinfo {author} {\bibfnamefont {T.~C.~H.}\ \bibnamefont
  {Liew}}, \bibinfo {author} {\bibfnamefont {P.~R.}\ \bibnamefont {Eastham}},
  \bibinfo {author} {\bibfnamefont {P.~G.}\ \bibnamefont {Savvidis}},\ and\
  \bibinfo {author} {\bibfnamefont {J.~J.}\ \bibnamefont {Baumberg}},\
  }\bibfield  {title} {\bibinfo {title} {Synchronization crossover of polariton
  condensates in weakly disordered lattices},\ }\href
  {https://doi.org/10.1103/PhysRevB.97.195109} {\bibfield  {journal} {\bibinfo
  {journal} {Phys. Rev. B}\ }\textbf {\bibinfo {volume} {97}},\ \bibinfo
  {pages} {195109} (\bibinfo {year} {2018})}\BibitemShut {NoStop}%
\bibitem [{\citenamefont {Klembt}\ \emph {et~al.}(2018)\citenamefont {Klembt},
  \citenamefont {Harder}, \citenamefont {Egorov}, \citenamefont {Winkler},
  \citenamefont {Ge}, \citenamefont {Bandres}, \citenamefont {Emmerling},
  \citenamefont {Worschech}, \citenamefont {Liew}, \citenamefont {Segev},
  \citenamefont {Schneider},\ and\ \citenamefont {H{\"o}fling}}]{Klembt2018}%
  \BibitemOpen
  \bibfield  {author} {\bibinfo {author} {\bibfnamefont {S.}~\bibnamefont
  {Klembt}}, \bibinfo {author} {\bibfnamefont {T.~H.}\ \bibnamefont {Harder}},
  \bibinfo {author} {\bibfnamefont {O.~A.}\ \bibnamefont {Egorov}}, \bibinfo
  {author} {\bibfnamefont {K.}~\bibnamefont {Winkler}}, \bibinfo {author}
  {\bibfnamefont {R.}~\bibnamefont {Ge}}, \bibinfo {author} {\bibfnamefont
  {M.~A.}\ \bibnamefont {Bandres}}, \bibinfo {author} {\bibfnamefont
  {M.}~\bibnamefont {Emmerling}}, \bibinfo {author} {\bibfnamefont
  {L.}~\bibnamefont {Worschech}}, \bibinfo {author} {\bibfnamefont {T.~C.~H.}\
  \bibnamefont {Liew}}, \bibinfo {author} {\bibfnamefont {M.}~\bibnamefont
  {Segev}}, \bibinfo {author} {\bibfnamefont {C.}~\bibnamefont {Schneider}},\
  and\ \bibinfo {author} {\bibfnamefont {S.}~\bibnamefont {H{\"o}fling}},\
  }\bibfield  {title} {\bibinfo {title} {Exciton-polariton topological
  insulator},\ }\href {https://doi.org/10.1038/s41586-018-0601-5} {\bibfield
  {journal} {\bibinfo  {journal} {Nature}\ }\textbf {\bibinfo {volume} {562}},\
  \bibinfo {pages} {552} (\bibinfo {year} {2018})}\BibitemShut {NoStop}%
\bibitem [{\citenamefont {Kartashov}\ and\ \citenamefont
  {Skryabin}(2019)}]{Kartashov2019}%
  \BibitemOpen
  \bibfield  {author} {\bibinfo {author} {\bibfnamefont {Y.~V.}\ \bibnamefont
  {Kartashov}}\ and\ \bibinfo {author} {\bibfnamefont {D.~V.}\ \bibnamefont
  {Skryabin}},\ }\bibfield  {title} {\bibinfo {title} {Two-dimensional
  topological polariton laser},\ }\href
  {https://doi.org/10.1103/PhysRevLett.122.083902} {\bibfield  {journal}
  {\bibinfo  {journal} {Phys. Rev. Lett.}\ }\textbf {\bibinfo {volume} {122}},\
  \bibinfo {pages} {083902} (\bibinfo {year} {2019})}\BibitemShut {NoStop}%
\bibitem [{\citenamefont {Sigurdsson}\ \emph {et~al.}(2019)\citenamefont
  {Sigurdsson}, \citenamefont {Krivosenko}, \citenamefont {Iorsh},
  \citenamefont {Shelykh},\ and\ \citenamefont {Nalitov}}]{Sigurdsson_PRB2019}%
  \BibitemOpen
  \bibfield  {author} {\bibinfo {author} {\bibfnamefont {H.}~\bibnamefont
  {Sigurdsson}}, \bibinfo {author} {\bibfnamefont {Y.~S.}\ \bibnamefont
  {Krivosenko}}, \bibinfo {author} {\bibfnamefont {I.~V.}\ \bibnamefont
  {Iorsh}}, \bibinfo {author} {\bibfnamefont {I.~A.}\ \bibnamefont {Shelykh}},\
  and\ \bibinfo {author} {\bibfnamefont {A.~V.}\ \bibnamefont {Nalitov}},\
  }\bibfield  {title} {\bibinfo {title} {Spontaneous topological transitions in
  a honeycomb lattice of exciton-polariton condensates due to spin
  bifurcations},\ }\href {https://doi.org/10.1103/PhysRevB.100.235444}
  {\bibfield  {journal} {\bibinfo  {journal} {Phys. Rev. B}\ }\textbf {\bibinfo
  {volume} {100}},\ \bibinfo {pages} {235444} (\bibinfo {year}
  {2019})}\BibitemShut {NoStop}%
\bibitem [{\citenamefont {Liu}\ \emph {et~al.}(2020)\citenamefont {Liu},
  \citenamefont {Ji}, \citenamefont {Wang}, \citenamefont {Modi}, \citenamefont
  {Hwang}, \citenamefont {Zheng}, \citenamefont {Sorger}, \citenamefont {Pan},\
  and\ \citenamefont {Agarwal}}]{Liu_Science2020}%
  \BibitemOpen
  \bibfield  {author} {\bibinfo {author} {\bibfnamefont {W.}~\bibnamefont
  {Liu}}, \bibinfo {author} {\bibfnamefont {Z.}~\bibnamefont {Ji}}, \bibinfo
  {author} {\bibfnamefont {Y.}~\bibnamefont {Wang}}, \bibinfo {author}
  {\bibfnamefont {G.}~\bibnamefont {Modi}}, \bibinfo {author} {\bibfnamefont
  {M.}~\bibnamefont {Hwang}}, \bibinfo {author} {\bibfnamefont
  {B.}~\bibnamefont {Zheng}}, \bibinfo {author} {\bibfnamefont {V.~J.}\
  \bibnamefont {Sorger}}, \bibinfo {author} {\bibfnamefont {A.}~\bibnamefont
  {Pan}},\ and\ \bibinfo {author} {\bibfnamefont {R.}~\bibnamefont {Agarwal}},\
  }\bibfield  {title} {\bibinfo {title} {Generation of helical topological
  exciton-polaritons},\ }\href {https://doi.org/10.1126/science.abc4975}
  {\bibfield  {journal} {\bibinfo  {journal} {Science}\ }\textbf {\bibinfo
  {volume} {370}},\ \bibinfo {pages} {600} (\bibinfo {year}
  {2020})}\BibitemShut {NoStop}%
\bibitem [{\citenamefont {Whittaker}\ \emph {et~al.}(2018)\citenamefont
  {Whittaker}, \citenamefont {Cancellieri}, \citenamefont {Walker},
  \citenamefont {Gulevich}, \citenamefont {Schomerus}, \citenamefont
  {Vaitiekus}, \citenamefont {Royall}, \citenamefont {Whittaker}, \citenamefont
  {Clarke}, \citenamefont {Iorsh}, \citenamefont {Shelykh}, \citenamefont
  {Skolnick},\ and\ \citenamefont {Krizhanovskii}}]{Whittaker2018}%
  \BibitemOpen
  \bibfield  {author} {\bibinfo {author} {\bibfnamefont {C.~E.}\ \bibnamefont
  {Whittaker}}, \bibinfo {author} {\bibfnamefont {E.}~\bibnamefont
  {Cancellieri}}, \bibinfo {author} {\bibfnamefont {P.~M.}\ \bibnamefont
  {Walker}}, \bibinfo {author} {\bibfnamefont {D.~R.}\ \bibnamefont
  {Gulevich}}, \bibinfo {author} {\bibfnamefont {H.}~\bibnamefont {Schomerus}},
  \bibinfo {author} {\bibfnamefont {D.}~\bibnamefont {Vaitiekus}}, \bibinfo
  {author} {\bibfnamefont {B.}~\bibnamefont {Royall}}, \bibinfo {author}
  {\bibfnamefont {D.~M.}\ \bibnamefont {Whittaker}}, \bibinfo {author}
  {\bibfnamefont {E.}~\bibnamefont {Clarke}}, \bibinfo {author} {\bibfnamefont
  {I.~V.}\ \bibnamefont {Iorsh}}, \bibinfo {author} {\bibfnamefont {I.~A.}\
  \bibnamefont {Shelykh}}, \bibinfo {author} {\bibfnamefont {M.~S.}\
  \bibnamefont {Skolnick}},\ and\ \bibinfo {author} {\bibfnamefont {D.~N.}\
  \bibnamefont {Krizhanovskii}},\ }\bibfield  {title} {\bibinfo {title}
  {Exciton polaritons in a two-dimensional lieb lattice with spin-orbit
  coupling},\ }\href {https://doi.org/10.1103/PhysRevLett.120.097401}
  {\bibfield  {journal} {\bibinfo  {journal} {Phys. Rev. Lett.}\ }\textbf
  {\bibinfo {volume} {120}},\ \bibinfo {pages} {097401} (\bibinfo {year}
  {2018})}\BibitemShut {NoStop}%
\bibitem [{\citenamefont {Goblot}\ \emph {et~al.}(2019)\citenamefont {Goblot},
  \citenamefont {Rauer}, \citenamefont {Vicentini}, \citenamefont {Le~Boit\'e},
  \citenamefont {Galopin}, \citenamefont {Lema\^{\i}tre}, \citenamefont
  {Le~Gratiet}, \citenamefont {Harouri}, \citenamefont {Sagnes}, \citenamefont
  {Ravets}, \citenamefont {Ciuti}, \citenamefont {Amo},\ and\ \citenamefont
  {Bloch}}]{Goblot2019}%
  \BibitemOpen
  \bibfield  {author} {\bibinfo {author} {\bibfnamefont {V.}~\bibnamefont
  {Goblot}}, \bibinfo {author} {\bibfnamefont {B.}~\bibnamefont {Rauer}},
  \bibinfo {author} {\bibfnamefont {F.}~\bibnamefont {Vicentini}}, \bibinfo
  {author} {\bibfnamefont {A.}~\bibnamefont {Le~Boit\'e}}, \bibinfo {author}
  {\bibfnamefont {E.}~\bibnamefont {Galopin}}, \bibinfo {author} {\bibfnamefont
  {A.}~\bibnamefont {Lema\^{\i}tre}}, \bibinfo {author} {\bibfnamefont
  {L.}~\bibnamefont {Le~Gratiet}}, \bibinfo {author} {\bibfnamefont
  {A.}~\bibnamefont {Harouri}}, \bibinfo {author} {\bibfnamefont
  {I.}~\bibnamefont {Sagnes}}, \bibinfo {author} {\bibfnamefont
  {S.}~\bibnamefont {Ravets}}, \bibinfo {author} {\bibfnamefont
  {C.}~\bibnamefont {Ciuti}}, \bibinfo {author} {\bibfnamefont
  {A.}~\bibnamefont {Amo}},\ and\ \bibinfo {author} {\bibfnamefont
  {J.}~\bibnamefont {Bloch}},\ }\bibfield  {title} {\bibinfo {title} {Nonlinear
  polariton fluids in a flatband reveal discrete gap solitons},\ }\href
  {https://doi.org/10.1103/PhysRevLett.123.113901} {\bibfield  {journal}
  {\bibinfo  {journal} {Phys. Rev. Lett.}\ }\textbf {\bibinfo {volume} {123}},\
  \bibinfo {pages} {113901} (\bibinfo {year} {2019})}\BibitemShut {NoStop}%
\bibitem [{\citenamefont {Berloff}\ \emph {et~al.}(2017)\citenamefont
  {Berloff}, \citenamefont {Silva}, \citenamefont {Kalinin}, \citenamefont
  {Askitopoulos}, \citenamefont {T{\"o}pfer}, \citenamefont {Cilibrizzi},
  \citenamefont {Langbein},\ and\ \citenamefont {Lagoudakis}}]{Berloff2017}%
  \BibitemOpen
  \bibfield  {author} {\bibinfo {author} {\bibfnamefont {N.~G.}\ \bibnamefont
  {Berloff}}, \bibinfo {author} {\bibfnamefont {M.}~\bibnamefont {Silva}},
  \bibinfo {author} {\bibfnamefont {K.}~\bibnamefont {Kalinin}}, \bibinfo
  {author} {\bibfnamefont {A.}~\bibnamefont {Askitopoulos}}, \bibinfo {author}
  {\bibfnamefont {J.~D.}\ \bibnamefont {T{\"o}pfer}}, \bibinfo {author}
  {\bibfnamefont {P.}~\bibnamefont {Cilibrizzi}}, \bibinfo {author}
  {\bibfnamefont {W.}~\bibnamefont {Langbein}},\ and\ \bibinfo {author}
  {\bibfnamefont {P.~G.}\ \bibnamefont {Lagoudakis}},\ }\bibfield  {title}
  {\bibinfo {title} {Realizing the classical xy hamiltonian in polariton
  simulators},\ }\href {https://doi.org/10.1038/nmat4971} {\bibfield  {journal}
  {\bibinfo  {journal} {Nature Materials}\ }\textbf {\bibinfo {volume} {16}},\
  \bibinfo {pages} {1120} (\bibinfo {year} {2017})}\BibitemShut {NoStop}%
\bibitem [{\citenamefont {Honari-Latifpour}\ and\ \citenamefont
  {Miri}(2020)}]{Miri2020}%
  \BibitemOpen
  \bibfield  {author} {\bibinfo {author} {\bibfnamefont {M.}~\bibnamefont
  {Honari-Latifpour}}\ and\ \bibinfo {author} {\bibfnamefont {M.-A.}\
  \bibnamefont {Miri}},\ }\bibfield  {title} {\bibinfo {title} {Mapping the
  $xy$ hamiltonian onto a network of coupled lasers},\ }\href
  {https://doi.org/10.1103/PhysRevResearch.2.043335} {\bibfield  {journal}
  {\bibinfo  {journal} {Phys. Rev. Research}\ }\textbf {\bibinfo {volume}
  {2}},\ \bibinfo {pages} {043335} (\bibinfo {year} {2020})}\BibitemShut
  {NoStop}%
\bibitem [{\citenamefont {Kalinin}\ and\ \citenamefont
  {Berloff}(2019)}]{Kalinin_PRB2019}%
  \BibitemOpen
  \bibfield  {author} {\bibinfo {author} {\bibfnamefont {K.~P.}\ \bibnamefont
  {Kalinin}}\ and\ \bibinfo {author} {\bibfnamefont {N.~G.}\ \bibnamefont
  {Berloff}},\ }\bibfield  {title} {\bibinfo {title} {Polaritonic network as a
  paradigm for dynamics of coupled oscillators},\ }\href
  {https://doi.org/10.1103/PhysRevB.100.245306} {\bibfield  {journal} {\bibinfo
   {journal} {Phys. Rev. B}\ }\textbf {\bibinfo {volume} {100}},\ \bibinfo
  {pages} {245306} (\bibinfo {year} {2019})}\BibitemShut {NoStop}%
\bibitem [{\citenamefont {Kalinin}\ and\ \citenamefont
  {Berloff}(2018{\natexlab{a}})}]{Kalinin_SciRep2018}%
  \BibitemOpen
  \bibfield  {author} {\bibinfo {author} {\bibfnamefont {K.~P.}\ \bibnamefont
  {Kalinin}}\ and\ \bibinfo {author} {\bibfnamefont {N.~G.}\ \bibnamefont
  {Berloff}},\ }\bibfield  {title} {\bibinfo {title} {Global optimization of
  spin hamiltonians with gain-dissipative systems},\ }\href
  {https://doi.org/10.1038/s41598-018-35416-1} {\bibfield  {journal} {\bibinfo
  {journal} {Scientific Reports}\ }\textbf {\bibinfo {volume} {8}},\ \bibinfo
  {pages} {17791} (\bibinfo {year} {2018}{\natexlab{a}})}\BibitemShut {NoStop}%
\bibitem [{\citenamefont {Kalinin}\ and\ \citenamefont
  {Berloff}(2018{\natexlab{b}})}]{Kalinin_PRL2018}%
  \BibitemOpen
  \bibfield  {author} {\bibinfo {author} {\bibfnamefont {K.~P.}\ \bibnamefont
  {Kalinin}}\ and\ \bibinfo {author} {\bibfnamefont {N.~G.}\ \bibnamefont
  {Berloff}},\ }\bibfield  {title} {\bibinfo {title} {Simulating ising and
  $n$-state planar potts models and external fields with nonequilibrium
  condensates},\ }\href {https://doi.org/10.1103/PhysRevLett.121.235302}
  {\bibfield  {journal} {\bibinfo  {journal} {Phys. Rev. Lett.}\ }\textbf
  {\bibinfo {volume} {121}},\ \bibinfo {pages} {235302} (\bibinfo {year}
  {2018}{\natexlab{b}})}\BibitemShut {NoStop}%
\bibitem [{\citenamefont {Kyriienko}\ \emph {et~al.}(2019)\citenamefont
  {Kyriienko}, \citenamefont {Sigurdsson},\ and\ \citenamefont
  {Liew}}]{Kyriienko_PRB2019}%
  \BibitemOpen
  \bibfield  {author} {\bibinfo {author} {\bibfnamefont {O.}~\bibnamefont
  {Kyriienko}}, \bibinfo {author} {\bibfnamefont {H.}~\bibnamefont
  {Sigurdsson}},\ and\ \bibinfo {author} {\bibfnamefont {T.~C.~H.}\
  \bibnamefont {Liew}},\ }\bibfield  {title} {\bibinfo {title} {Probabilistic
  solving of $np$-hard problems with bistable nonlinear optical networks},\
  }\href {https://doi.org/10.1103/PhysRevB.99.195301} {\bibfield  {journal}
  {\bibinfo  {journal} {Phys. Rev. B}\ }\textbf {\bibinfo {volume} {99}},\
  \bibinfo {pages} {195301} (\bibinfo {year} {2019})}\BibitemShut {NoStop}%
\bibitem [{\citenamefont {Baboux}\ \emph {et~al.}(2018)\citenamefont {Baboux},
  \citenamefont {Bernardis}, \citenamefont {Goblot}, \citenamefont {Gladilin},
  \citenamefont {Gomez}, \citenamefont {Galopin}, \citenamefont {Gratiet},
  \citenamefont {Lema\^{i}tre}, \citenamefont {Sagnes}, \citenamefont
  {Carusotto}, \citenamefont {Wouters}, \citenamefont {Amo},\ and\
  \citenamefont {Bloch}}]{Baboux_Optica2018}%
  \BibitemOpen
  \bibfield  {author} {\bibinfo {author} {\bibfnamefont {F.}~\bibnamefont
  {Baboux}}, \bibinfo {author} {\bibfnamefont {D.~D.}\ \bibnamefont
  {Bernardis}}, \bibinfo {author} {\bibfnamefont {V.}~\bibnamefont {Goblot}},
  \bibinfo {author} {\bibfnamefont {V.~N.}\ \bibnamefont {Gladilin}}, \bibinfo
  {author} {\bibfnamefont {C.}~\bibnamefont {Gomez}}, \bibinfo {author}
  {\bibfnamefont {E.}~\bibnamefont {Galopin}}, \bibinfo {author} {\bibfnamefont
  {L.~L.}\ \bibnamefont {Gratiet}}, \bibinfo {author} {\bibfnamefont
  {A.}~\bibnamefont {Lema\^{i}tre}}, \bibinfo {author} {\bibfnamefont
  {I.}~\bibnamefont {Sagnes}}, \bibinfo {author} {\bibfnamefont
  {I.}~\bibnamefont {Carusotto}}, \bibinfo {author} {\bibfnamefont
  {M.}~\bibnamefont {Wouters}}, \bibinfo {author} {\bibfnamefont
  {A.}~\bibnamefont {Amo}},\ and\ \bibinfo {author} {\bibfnamefont
  {J.}~\bibnamefont {Bloch}},\ }\bibfield  {title} {\bibinfo {title} {Unstable
  and stable regimes of polariton condensation},\ }\href
  {https://doi.org/10.1364/OPTICA.5.001163} {\bibfield  {journal} {\bibinfo
  {journal} {Optica}\ }\textbf {\bibinfo {volume} {5}},\ \bibinfo {pages}
  {1163} (\bibinfo {year} {2018})}\BibitemShut {NoStop}%
\bibitem [{\citenamefont {Liew}\ \emph {et~al.}(2015)\citenamefont {Liew},
  \citenamefont {Egorov}, \citenamefont {Matuszewski}, \citenamefont
  {Kyriienko}, \citenamefont {Ma},\ and\ \citenamefont
  {Ostrovskaya}}]{Liew_PRB2015}%
  \BibitemOpen
  \bibfield  {author} {\bibinfo {author} {\bibfnamefont {T.~C.~H.}\
  \bibnamefont {Liew}}, \bibinfo {author} {\bibfnamefont {O.~A.}\ \bibnamefont
  {Egorov}}, \bibinfo {author} {\bibfnamefont {M.}~\bibnamefont {Matuszewski}},
  \bibinfo {author} {\bibfnamefont {O.}~\bibnamefont {Kyriienko}}, \bibinfo
  {author} {\bibfnamefont {X.}~\bibnamefont {Ma}},\ and\ \bibinfo {author}
  {\bibfnamefont {E.~A.}\ \bibnamefont {Ostrovskaya}},\ }\bibfield  {title}
  {\bibinfo {title} {Instability-induced formation and nonequilibrium dynamics
  of phase defects in polariton condensates},\ }\href
  {https://doi.org/10.1103/PhysRevB.91.085413} {\bibfield  {journal} {\bibinfo
  {journal} {Phys. Rev. B}\ }\textbf {\bibinfo {volume} {91}},\ \bibinfo
  {pages} {085413} (\bibinfo {year} {2015})}\BibitemShut {NoStop}%
\bibitem [{\citenamefont {Werner}\ \emph {et~al.}(2014)\citenamefont {Werner},
  \citenamefont {Egorov},\ and\ \citenamefont {Lederer}}]{Werner_PRB2014}%
  \BibitemOpen
  \bibfield  {author} {\bibinfo {author} {\bibfnamefont {A.}~\bibnamefont
  {Werner}}, \bibinfo {author} {\bibfnamefont {O.~A.}\ \bibnamefont {Egorov}},\
  and\ \bibinfo {author} {\bibfnamefont {F.}~\bibnamefont {Lederer}},\
  }\bibfield  {title} {\bibinfo {title} {Exciton-polariton patterns in
  coherently pumped semiconductor microcavities},\ }\href
  {https://doi.org/10.1103/PhysRevB.89.245307} {\bibfield  {journal} {\bibinfo
  {journal} {Phys. Rev. B}\ }\textbf {\bibinfo {volume} {89}},\ \bibinfo
  {pages} {245307} (\bibinfo {year} {2014})}\BibitemShut {NoStop}%
\bibitem [{\citenamefont {Dagvadorj}\ \emph {et~al.}(2015)\citenamefont
  {Dagvadorj}, \citenamefont {Fellows}, \citenamefont
  {Matyja\ifmmode~\acute{s}\else \'{s}\fi{}kiewicz}, \citenamefont {Marchetti},
  \citenamefont {Carusotto},\ and\ \citenamefont {Szyma\ifmmode~\acute{n}\else
  \'{n}\fi{}ska}}]{Dagvadorj_PRX2015}%
  \BibitemOpen
  \bibfield  {author} {\bibinfo {author} {\bibfnamefont {G.}~\bibnamefont
  {Dagvadorj}}, \bibinfo {author} {\bibfnamefont {J.~M.}\ \bibnamefont
  {Fellows}}, \bibinfo {author} {\bibfnamefont {S.}~\bibnamefont
  {Matyja\ifmmode~\acute{s}\else \'{s}\fi{}kiewicz}}, \bibinfo {author}
  {\bibfnamefont {F.~M.}\ \bibnamefont {Marchetti}}, \bibinfo {author}
  {\bibfnamefont {I.}~\bibnamefont {Carusotto}},\ and\ \bibinfo {author}
  {\bibfnamefont {M.~H.}\ \bibnamefont {Szyma\ifmmode~\acute{n}\else
  \'{n}\fi{}ska}},\ }\bibfield  {title} {\bibinfo {title} {Nonequilibrium phase
  transition in a two-dimensional driven open quantum system},\ }\href
  {https://doi.org/10.1103/PhysRevX.5.041028} {\bibfield  {journal} {\bibinfo
  {journal} {Phys. Rev. X}\ }\textbf {\bibinfo {volume} {5}},\ \bibinfo {pages}
  {041028} (\bibinfo {year} {2015})}\BibitemShut {NoStop}%
\bibitem [{\citenamefont {Zamora}\ \emph {et~al.}(2020)\citenamefont {Zamora},
  \citenamefont {Dagvadorj}, \citenamefont {Comaron}, \citenamefont
  {Carusotto}, \citenamefont {Proukakis},\ and\ \citenamefont
  {Szyma\ifmmode~\acute{n}\else \'{n}\fi{}ska}}]{Zamore_PRL2020}%
  \BibitemOpen
  \bibfield  {author} {\bibinfo {author} {\bibfnamefont {A.}~\bibnamefont
  {Zamora}}, \bibinfo {author} {\bibfnamefont {G.}~\bibnamefont {Dagvadorj}},
  \bibinfo {author} {\bibfnamefont {P.}~\bibnamefont {Comaron}}, \bibinfo
  {author} {\bibfnamefont {I.}~\bibnamefont {Carusotto}}, \bibinfo {author}
  {\bibfnamefont {N.~P.}\ \bibnamefont {Proukakis}},\ and\ \bibinfo {author}
  {\bibfnamefont {M.~H.}\ \bibnamefont {Szyma\ifmmode~\acute{n}\else
  \'{n}\fi{}ska}},\ }\bibfield  {title} {\bibinfo {title} {Kibble-zurek
  mechanism in driven dissipative systems crossing a nonequilibrium phase
  transition},\ }\href {https://doi.org/10.1103/PhysRevLett.125.095301}
  {\bibfield  {journal} {\bibinfo  {journal} {Phys. Rev. Lett.}\ }\textbf
  {\bibinfo {volume} {125}},\ \bibinfo {pages} {095301} (\bibinfo {year}
  {2020})}\BibitemShut {NoStop}%
\bibitem [{\citenamefont {Comaron}\ \emph {et~al.}(2018)\citenamefont
  {Comaron}, \citenamefont {Dagvadorj}, \citenamefont {Zamora}, \citenamefont
  {Carusotto}, \citenamefont {Proukakis},\ and\ \citenamefont
  {Szyma\ifmmode~\acute{n}\else \'{n}\fi{}ska}}]{Comaron_PRL2018}%
  \BibitemOpen
  \bibfield  {author} {\bibinfo {author} {\bibfnamefont {P.}~\bibnamefont
  {Comaron}}, \bibinfo {author} {\bibfnamefont {G.}~\bibnamefont {Dagvadorj}},
  \bibinfo {author} {\bibfnamefont {A.}~\bibnamefont {Zamora}}, \bibinfo
  {author} {\bibfnamefont {I.}~\bibnamefont {Carusotto}}, \bibinfo {author}
  {\bibfnamefont {N.~P.}\ \bibnamefont {Proukakis}},\ and\ \bibinfo {author}
  {\bibfnamefont {M.~H.}\ \bibnamefont {Szyma\ifmmode~\acute{n}\else
  \'{n}\fi{}ska}},\ }\bibfield  {title} {\bibinfo {title} {Dynamical critical
  exponents in driven-dissipative quantum systems},\ }\href
  {https://doi.org/10.1103/PhysRevLett.121.095302} {\bibfield  {journal}
  {\bibinfo  {journal} {Phys. Rev. Lett.}\ }\textbf {\bibinfo {volume} {121}},\
  \bibinfo {pages} {095302} (\bibinfo {year} {2018})}\BibitemShut {NoStop}%
\bibitem [{\citenamefont {Solnyshkov}\ \emph {et~al.}(2016)\citenamefont
  {Solnyshkov}, \citenamefont {Nalitov},\ and\ \citenamefont
  {Malpuech}}]{Solnyshkov_PRL2016}%
  \BibitemOpen
  \bibfield  {author} {\bibinfo {author} {\bibfnamefont {D.~D.}\ \bibnamefont
  {Solnyshkov}}, \bibinfo {author} {\bibfnamefont {A.~V.}\ \bibnamefont
  {Nalitov}},\ and\ \bibinfo {author} {\bibfnamefont {G.}~\bibnamefont
  {Malpuech}},\ }\bibfield  {title} {\bibinfo {title} {Kibble-zurek mechanism
  in topologically nontrivial zigzag chains of polariton micropillars},\ }\href
  {https://doi.org/10.1103/PhysRevLett.116.046402} {\bibfield  {journal}
  {\bibinfo  {journal} {Phys. Rev. Lett.}\ }\textbf {\bibinfo {volume} {116}},\
  \bibinfo {pages} {046402} (\bibinfo {year} {2016})}\BibitemShut {NoStop}%
\bibitem [{\citenamefont {Wilson}\ \emph {et~al.}(2016)\citenamefont {Wilson},
  \citenamefont {Mahmud}, \citenamefont {Hu}, \citenamefont {Gorshkov},
  \citenamefont {Hafezi},\ and\ \citenamefont {Foss-Feig}}]{Wilson2016}%
  \BibitemOpen
  \bibfield  {author} {\bibinfo {author} {\bibfnamefont {R.~M.}\ \bibnamefont
  {Wilson}}, \bibinfo {author} {\bibfnamefont {K.~W.}\ \bibnamefont {Mahmud}},
  \bibinfo {author} {\bibfnamefont {A.}~\bibnamefont {Hu}}, \bibinfo {author}
  {\bibfnamefont {A.~V.}\ \bibnamefont {Gorshkov}}, \bibinfo {author}
  {\bibfnamefont {M.}~\bibnamefont {Hafezi}},\ and\ \bibinfo {author}
  {\bibfnamefont {M.}~\bibnamefont {Foss-Feig}},\ }\bibfield  {title} {\bibinfo
  {title} {Collective phases of strongly interacting cavity photons},\ }\href
  {https://doi.org/10.1103/PhysRevA.94.033801} {\bibfield  {journal} {\bibinfo
  {journal} {Phys. Rev. A}\ }\textbf {\bibinfo {volume} {94}},\ \bibinfo
  {pages} {033801} (\bibinfo {year} {2016})}\BibitemShut {NoStop}%
\bibitem [{\citenamefont {Koniakhin}\ \emph {et~al.}(2019)\citenamefont
  {Koniakhin}, \citenamefont {Bleu}, \citenamefont {Stupin}, \citenamefont
  {Pigeon}, \citenamefont {Maitre}, \citenamefont {Claude}, \citenamefont
  {Lerario}, \citenamefont {Glorieux}, \citenamefont {Bramati}, \citenamefont
  {Solnyshkov},\ and\ \citenamefont {Malpuech}}]{Koniakhin_PRL2019}%
  \BibitemOpen
  \bibfield  {author} {\bibinfo {author} {\bibfnamefont {S.~V.}\ \bibnamefont
  {Koniakhin}}, \bibinfo {author} {\bibfnamefont {O.}~\bibnamefont {Bleu}},
  \bibinfo {author} {\bibfnamefont {D.~D.}\ \bibnamefont {Stupin}}, \bibinfo
  {author} {\bibfnamefont {S.}~\bibnamefont {Pigeon}}, \bibinfo {author}
  {\bibfnamefont {A.}~\bibnamefont {Maitre}}, \bibinfo {author} {\bibfnamefont
  {F.}~\bibnamefont {Claude}}, \bibinfo {author} {\bibfnamefont
  {G.}~\bibnamefont {Lerario}}, \bibinfo {author} {\bibfnamefont
  {Q.}~\bibnamefont {Glorieux}}, \bibinfo {author} {\bibfnamefont
  {A.}~\bibnamefont {Bramati}}, \bibinfo {author} {\bibfnamefont
  {D.}~\bibnamefont {Solnyshkov}},\ and\ \bibinfo {author} {\bibfnamefont
  {G.}~\bibnamefont {Malpuech}},\ }\bibfield  {title} {\bibinfo {title}
  {Stationary quantum vortex street in a driven-dissipative quantum fluid of
  light},\ }\href {https://doi.org/10.1103/PhysRevLett.123.215301} {\bibfield
  {journal} {\bibinfo  {journal} {Phys. Rev. Lett.}\ }\textbf {\bibinfo
  {volume} {123}},\ \bibinfo {pages} {215301} (\bibinfo {year}
  {2019})}\BibitemShut {NoStop}%
\bibitem [{\citenamefont {Sigurdsson}\ \emph
  {et~al.}(2017{\natexlab{a}})\citenamefont {Sigurdsson}, \citenamefont
  {Liew},\ and\ \citenamefont {Shelykh}}]{Sigurdsson_PRB2017_parsol}%
  \BibitemOpen
  \bibfield  {author} {\bibinfo {author} {\bibfnamefont {H.}~\bibnamefont
  {Sigurdsson}}, \bibinfo {author} {\bibfnamefont {T.~C.~H.}\ \bibnamefont
  {Liew}},\ and\ \bibinfo {author} {\bibfnamefont {I.~A.}\ \bibnamefont
  {Shelykh}},\ }\bibfield  {title} {\bibinfo {title} {Parity solitons in
  nonresonantly driven-dissipative condensate channels},\ }\href
  {https://doi.org/10.1103/PhysRevB.96.205406} {\bibfield  {journal} {\bibinfo
  {journal} {Phys. Rev. B}\ }\textbf {\bibinfo {volume} {96}},\ \bibinfo
  {pages} {205406} (\bibinfo {year} {2017}{\natexlab{a}})}\BibitemShut
  {NoStop}%
\bibitem [{\citenamefont {Sigurdsson}\ \emph
  {et~al.}(2017{\natexlab{b}})\citenamefont {Sigurdsson}, \citenamefont
  {Ramsay}, \citenamefont {Ohadi}, \citenamefont {Rubo}, \citenamefont {Liew},
  \citenamefont {Baumberg},\ and\ \citenamefont {Shelykh}}]{Sigurdsson2017a}%
  \BibitemOpen
  \bibfield  {author} {\bibinfo {author} {\bibfnamefont {H.}~\bibnamefont
  {Sigurdsson}}, \bibinfo {author} {\bibfnamefont {A.~J.}\ \bibnamefont
  {Ramsay}}, \bibinfo {author} {\bibfnamefont {H.}~\bibnamefont {Ohadi}},
  \bibinfo {author} {\bibfnamefont {Y.~G.}\ \bibnamefont {Rubo}}, \bibinfo
  {author} {\bibfnamefont {T.~C.~H.}\ \bibnamefont {Liew}}, \bibinfo {author}
  {\bibfnamefont {J.~J.}\ \bibnamefont {Baumberg}},\ and\ \bibinfo {author}
  {\bibfnamefont {I.~A.}\ \bibnamefont {Shelykh}},\ }\bibfield  {title}
  {\bibinfo {title} {Driven-dissipative spin chain model based on
  exciton-polariton condensates},\ }\href
  {https://doi.org/10.1103/PhysRevB.96.155403} {\bibfield  {journal} {\bibinfo
  {journal} {Phys. Rev. B}\ }\textbf {\bibinfo {volume} {96}},\ \bibinfo
  {pages} {155403} (\bibinfo {year} {2017}{\natexlab{b}})}\BibitemShut
  {NoStop}%
\bibitem [{\citenamefont {Ohadi}\ \emph {et~al.}(2015)\citenamefont {Ohadi},
  \citenamefont {Dreismann}, \citenamefont {Rubo}, \citenamefont {Pinsker},
  \citenamefont {del Valle-Inclan~Redondo}, \citenamefont {Tsintzos},
  \citenamefont {Hatzopoulos}, \citenamefont {Savvidis},\ and\ \citenamefont
  {Baumberg}}]{Ohadi2015}%
  \BibitemOpen
  \bibfield  {author} {\bibinfo {author} {\bibfnamefont {H.}~\bibnamefont
  {Ohadi}}, \bibinfo {author} {\bibfnamefont {A.}~\bibnamefont {Dreismann}},
  \bibinfo {author} {\bibfnamefont {Y.~G.}\ \bibnamefont {Rubo}}, \bibinfo
  {author} {\bibfnamefont {F.}~\bibnamefont {Pinsker}}, \bibinfo {author}
  {\bibfnamefont {Y.}~\bibnamefont {del Valle-Inclan~Redondo}}, \bibinfo
  {author} {\bibfnamefont {S.~I.}\ \bibnamefont {Tsintzos}}, \bibinfo {author}
  {\bibfnamefont {Z.}~\bibnamefont {Hatzopoulos}}, \bibinfo {author}
  {\bibfnamefont {P.~G.}\ \bibnamefont {Savvidis}},\ and\ \bibinfo {author}
  {\bibfnamefont {J.~J.}\ \bibnamefont {Baumberg}},\ }\bibfield  {title}
  {\bibinfo {title} {Spontaneous spin bifurcations and ferromagnetic phase
  transitions in a spinor exciton-polariton condensate},\ }\href
  {https://doi.org/10.1103/PhysRevX.5.031002} {\bibfield  {journal} {\bibinfo
  {journal} {Phys. Rev. X}\ }\textbf {\bibinfo {volume} {5}},\ \bibinfo {pages}
  {031002} (\bibinfo {year} {2015})}\BibitemShut {NoStop}%
\bibitem [{\citenamefont {Read}\ \emph {et~al.}(2009)\citenamefont {Read},
  \citenamefont {Liew}, \citenamefont {Rubo},\ and\ \citenamefont
  {Kavokin}}]{Read_PRB2009}%
  \BibitemOpen
  \bibfield  {author} {\bibinfo {author} {\bibfnamefont {D.}~\bibnamefont
  {Read}}, \bibinfo {author} {\bibfnamefont {T.~C.~H.}\ \bibnamefont {Liew}},
  \bibinfo {author} {\bibfnamefont {Y.~G.}\ \bibnamefont {Rubo}},\ and\
  \bibinfo {author} {\bibfnamefont {A.~V.}\ \bibnamefont {Kavokin}},\
  }\bibfield  {title} {\bibinfo {title} {Stochastic polarization formation in
  exciton-polariton bose-einstein condensates},\ }\href
  {https://doi.org/10.1103/PhysRevB.80.195309} {\bibfield  {journal} {\bibinfo
  {journal} {Phys. Rev. B}\ }\textbf {\bibinfo {volume} {80}},\ \bibinfo
  {pages} {195309} (\bibinfo {year} {2009})}\BibitemShut {NoStop}%
\bibitem [{\citenamefont {Dreismann}\ \emph {et~al.}(2016)\citenamefont
  {Dreismann}, \citenamefont {Ohadi}, \citenamefont {del Valle-Inclan~Redondo},
  \citenamefont {Balili}, \citenamefont {Rubo}, \citenamefont {Tsintzos},
  \citenamefont {Deligeorgis}, \citenamefont {Hatzopoulos}, \citenamefont
  {Savvidis},\ and\ \citenamefont {Baumberg}}]{Dreismann_NatMat2016}%
  \BibitemOpen
  \bibfield  {author} {\bibinfo {author} {\bibfnamefont {A.}~\bibnamefont
  {Dreismann}}, \bibinfo {author} {\bibfnamefont {H.}~\bibnamefont {Ohadi}},
  \bibinfo {author} {\bibfnamefont {Y.}~\bibnamefont {del
  Valle-Inclan~Redondo}}, \bibinfo {author} {\bibfnamefont {R.}~\bibnamefont
  {Balili}}, \bibinfo {author} {\bibfnamefont {Y.~G.}\ \bibnamefont {Rubo}},
  \bibinfo {author} {\bibfnamefont {S.~I.}\ \bibnamefont {Tsintzos}}, \bibinfo
  {author} {\bibfnamefont {G.}~\bibnamefont {Deligeorgis}}, \bibinfo {author}
  {\bibfnamefont {Z.}~\bibnamefont {Hatzopoulos}}, \bibinfo {author}
  {\bibfnamefont {P.~G.}\ \bibnamefont {Savvidis}},\ and\ \bibinfo {author}
  {\bibfnamefont {J.~J.}\ \bibnamefont {Baumberg}},\ }\bibfield  {title}
  {\bibinfo {title} {A sub-femtojoule electrical spin-switch based on optically
  trapped polariton condensates},\ }\href {https://doi.org/10.1038/nmat4722}
  {\bibfield  {journal} {\bibinfo  {journal} {Nature Materials}\ }\textbf
  {\bibinfo {volume} {15}},\ \bibinfo {pages} {1074} (\bibinfo {year}
  {2016})}\BibitemShut {NoStop}%
\bibitem [{\citenamefont {van~der Maaten}\ and\ \citenamefont
  {Hinton}(2008)}]{vanderMaaten2008}%
  \BibitemOpen
  \bibfield  {author} {\bibinfo {author} {\bibfnamefont {L.}~\bibnamefont
  {van~der Maaten}}\ and\ \bibinfo {author} {\bibfnamefont {G.}~\bibnamefont
  {Hinton}},\ }\bibfield  {title} {\bibinfo {title} {Visualizing data using
  t-sne},\ }\href {http://jmlr.org/papers/v9/vandermaaten08a.html} {\bibfield
  {journal} {\bibinfo  {journal} {Journal of Machine Learning Research}\
  }\textbf {\bibinfo {volume} {9}},\ \bibinfo {pages} {2579} (\bibinfo {year}
  {2008})}\BibitemShut {NoStop}%
\end{thebibliography}
\end{document}